\documentclass[ reprint,
 superscriptaddress,
 amsmath,
 amssymb,
 aip,
 a4paper,
 floatfix,
 longbibliography,
 twocolumn]{revtex4-2}

\usepackage{graphicx}
\usepackage{dcolumn}
\usepackage{bm}

\usepackage{physics}

\usepackage{xcolor}

\usepackage{microtype}

\usepackage[american]{babel}
\usepackage[utf8x]{inputenc}

\usepackage{siunitx} 
\usepackage{booktabs} 
\usepackage{multirow}

\newcommand{\ee}{\text{e}}
\newcommand{\ii}{\text{i}}

\newcommand{\tauzero}{\tau_{0}}
\newcommand{\abstau}{\abs{\tau}}
\newcommand{\limtauzeroinfty}{\lim_{\tauzero\rightarrow\infty}}

\newcommand{\fast}{f^{\ast}}

\newcommand{\zsquard}{z^{2}}
\newcommand{\zsquardzero}{z_{0}^{2}}
\newcommand{\zzero}{z_{0}}
\newcommand{\mathcalZzero}{\mathcal{Z}_{0}}

\newcommand{\fabsminustau}{f(-\abs{\tau})}
\newcommand{\fabsplustau}{f(\abs{\tau})}

\newcommand{\fastabsminustau}{f^{\ast}(-\abs{\tau})}
\newcommand{\fastabsplustau}{f^{\ast}(\abs{\tau})}

\newcommand{\fminustauzero}{f(-\tau_{0})}
\newcommand{\fastminustauzero}{f^{\ast}(-\tau_{0})}

\newcommand{\fplustauzero}{f(\tauzero)}
\newcommand{\fastplustauzero}{f^{\ast}(\tauzero)}

\newcommand{\parabolicfunction}[2]{\mathcal{D}_{#1}(#2)}

\newcommand{\orcid}[1]{\href{https://orcid.org/#1}{\includegraphics[width=7pt]{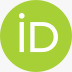}}}

\definecolor{cset-aps-blueberry}{RGB}{28,128,158}
\definecolor{cset-aps-blue}{RGB}{46,44,184}
\definecolor{cset-aps-turquoise}{RGB}{0,67,88}
\definecolor{cset-aps-limegreen}{RGB}{190,219,67}
\definecolor{cset-aps-green}{RGB}{31,138,112}
\definecolor{cset-aps-yellow}{RGB}{255,225,25}
\definecolor{cset-aps-orange}{RGB}{253,116,0}
\definecolor{cset-aps-red}{RGB}{219,0,43}
\definecolor{myred}{RGB}{255,0,20}

\usepackage{hyperref}
\hypersetup{%
    colorlinks=true,
    linkcolor={cset-aps-red},
    linkbordercolor={cset-aps-red},
    filecolor={cset-aps-orange},
    filebordercolor={cset-aps-orange},
    citecolor={cset-aps-blue},
    citebordercolor={cset-aps-blue},
    urlcolor={cset-aps-blueberry},
    urlbordercolor={cset-aps-blueberry},
    menucolor={cset-aps-limegreen},
    menubordercolor={cset-aps-limegreen},
    breaklinks=true,
    pdfborderstyle={/S/U/W 2},
    pdfpagemode=UseOutlines,
    pdfstartpage={1},
}

\usepackage{cleveref}

\AtBeginDocument{\RenewCommandCopy\qty\SI}

\begin{document}

\preprint{APS/123-QED}

\title{Elementary asymptotic approach to the Landau-Zener problem}

\author{Eric P. Glasbrenner \orcid{0000-0002-0822-3888}}
\affiliation{Institut f{\"u}r Quantenphysik and Center for Integrated Quantum
    Science and Technology (IQST), Universit{\"a}t Ulm, Albert-Einstein-Allee 11, D-89081 Ulm, Germany}
\email{eric.glasbrenner@uni-ulm.de}

\author{Wolfgang P. Schleich \orcid{0000-0002-9693-8882}}
\affiliation{Institut f{\"u}r Quantenphysik and Center for Integrated Quantum
    Science and Technology (IQST), Universit{\"a}t Ulm, Albert-Einstein-Allee 11, D-89081 Ulm, Germany}%
\affiliation{Institute for Quantum Science and Engineering (IQSE), and Texas A\&M AgriLife Research and Hagler Institute for Advanced Study, Texas A\&M University, College Station, TX 77843-4242, USA}

\date{\today}

\begin{abstract}
We present an asymptotic approach towards the standard Landau-Zener problem based on two linearly independent elementary waves of constant amplitude but time-dependent phase.
The two contributions to this phase are quadratic and logarithmic in time and result from the linear chirp of the energies and the lowest order correction in the coupling between the two levels in the long-time limit.
Indeed, our solutions subjected to initial conditions at a large but finite time in the past, are valid for large negative and large positive times. 
Due to their asymptotic nature they are not valid in the neighborhood of the moment when the levels cross.
However, as the starting point of the dynamics moves further into the past, the time interval of the break-down of our asymptotic solutions shrinks and vanishes in the limit of the infinite past which corresponds to the standard Landau-Zener situation.
Our approach explains not only every feature of the exact solution but yields deeper insights into the origin of the effects.
In particular, it (i) brings to light the subtleties involved in the asymptotic limit leading to the standard expressions for the Landau-Zener transition amplitudes, (ii) identifies the logarithmic phase as the origin of the exponential transition probability amplitude, and (iii) reveals the structure of the lowest order corrections to the Landau-Zener result when the starting point is not in the infinite past.
\end{abstract}

\maketitle

\section{Introduction}
\label{section:introduction}

A powerful technique to solve a differential equation relies on its asymptotic behavior, that is, to analyze it in a domain where elementary solutions exist.
However, the main problem of this approach is to merge the solutions in the different limits \cite{Berry1972, bender1999advanced}.
In the present article, we apply this technique to bring light into characteristic features of the Landau-Zener effect \cite{Landau1932a, Landau1932b, Zener1932, Majorana1932, Stueckelberg1932}, and demonstrate that an inherent logarithmic phase singularity \cite{Ullinger2022, Glasbrenner2023, Glasbrenner2024book, Glasbrenner2025} allows us to match the solutions in the two asymptotic regimes of large negative and large positive times.

The Landau-Zener effect stands as a paradigm of time-dependent quantum dynamics.
Describing the non-adiabatic transition \cite{Dykhne1960, Dykhne1962, Davis1976, Berry1990, nakamura2012} between two energy levels at an avoided crossing, it is central to numerous branches of physics \cite{Hill1953, bjoerkman2025, Wubs2005, Wubs2006, Saito2007, Peik1997, Oliver2005, Morsch2006, Dziarmaga2005, Lim2025}.
More recently, it has found renewed relevance in quantum control \cite{Ribeiro2009, SHEVCHENKO2010, Vitanov2017, ivakhnenko2023, kofman2024} and atom interferometry \cite{Siemss2020, Pagel2020, Gebbe2021, Fitzek2024, Konrad2024}.

Due to its prominence, the Landau-Zener problem has been revisited frequently, and a 'quasi-infinite' number of derivations exist in the literature \cite{Wittig2005, Vutha2010, Sun2025}.
These typically employ sophisticated mathematical tools, ranging from the exact solution in terms of parabolic cylinder functions to complex WKB analyses \cite{Zhu1992, Vitanov1996, Teranishi1997, Vitanov1999, Matsuda2025, Berry1993, Betz2006, Kofman2023, DeGrandi2010, Kayanuma1997}.
However, these approaches share one common drawback, that is, the physical origin of the transition probability is often rather opaque, hidden behind a curtain of special functions or complex contour integrals.

In the present article, we introduce a rather elementary approach to this cornerstone of quantum mechanics based on two linearly independent approximate but analytical solutions valid in the limit of large negative and large positive times.
The corresponding asymptotic waves have a phase which depends logarithmically on time.
It is the analytical continuation of this logarithm from positive to negative times which contains the essence of the Landau-Zener transition giving rise to a factor $\ii\pi$, and thus to the familiar Landau-Zener probability amplitude \cite{Glasbrenner2023, Glasbrenner2024book, Glasbrenner2025}. 

Our article is organized as follows:
In order to set the stage we first summarize in \cref{section:2} the key features of the Landau-Zener model. We then present in \cref{section:3} approximate but analytical expressions for the two probability amplitudes in terms of superpositions of two elementary waves valid in the limit of large negative as well as large positive times and obtain the dynamics of the Landau-Zener problem.
In this process we identify the logarithmic phase singularity contained in the elementary waves as the origin of the well-known Landau-Zener transition probability.
In \cref{section:5}, we conclude by briefly summarizing our results and providing an outlook. 

In order to keep the article self-contained yet concise, we have included two appendices.
We devote \mbox{Appendix \ref{appendix:B}} to a derivation of our asymptotic solutions starting from the exact expressions for the two probability amplitudes in terms of parabolic cylinder functions.
Here we concentrate on large but finite values of the initial time.
Moreover, in \mbox{Appendix \ref{appendix:A}} we obtain our two linearly independent elementary waves from the WKB-approximation of the Schrödinger-like differential equation of second order for one of the probability amplitudes.


\section{Landau-Zener in a Nutshell}
\label{section:2}

In this section, we first recall the differential equations governing the Landau-Zener problem and then summarize their key predictions by numerically solving them.
Throughout this section, we do not present any explanations but only describe the emerging phenomena. 
Our asymptotic approach described in detail in the following section will then provide the missing insights. 

\begin{figure}[htbp]
    \includegraphics[width=\columnwidth]{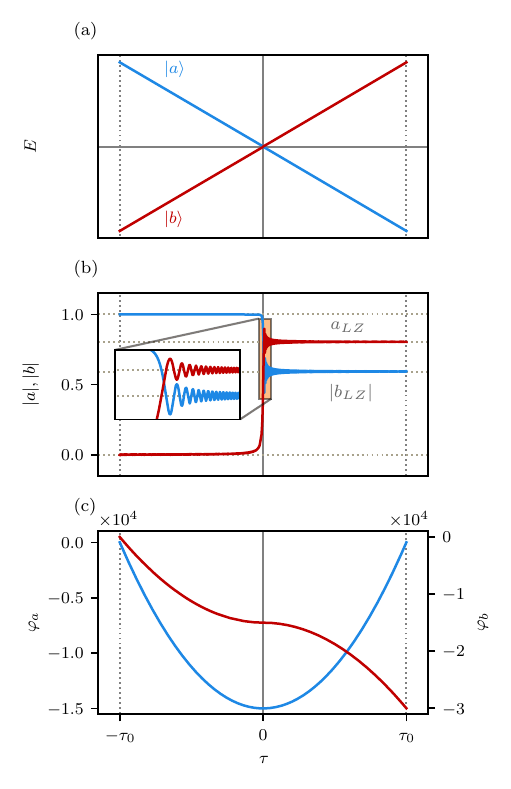}
    \centering
    \caption[]{Quantum dynamics in the Landau-Zener level-crossing problem.
    At $\tau = -\tauzero$ the two energy levels denoted by $\ket{a}$ and $\ket{b}$ (a) start to approach each other linearly in time and cross at $\tau=0$.
    We address the question of the values of the corresponding probability amplitudes $a$ and $b$ at $\tau = \tauzero$ when $\tauzero\rightarrow\infty$.
    In panel (b), we present the time evolution of the absolute values $\abs{a}$ and $ \abs{b} $ and in (c) the phases $\varphi_a$ and $\varphi_b$ of the two probability amplitudes $ a = a(\tau) $ and $ b = b(\tau) $. 
    We emphasize the sudden transitions of $\abs{a}$ and $\abs{b}$ in the vicinity of $\tau=0$ from $\abs{a(-\tauzero)} = 1$ and \mbox{$\abs{b(-\tauzero)}=0$} to \mbox{$\lim_{\tauzero \to \infty} \abs{a(\tauzero)} = \exp(-\pi/(2\epsilon))$} and \mbox{$\lim_{\tauzero \to \infty} \abs{b(\tauzero)} = \sqrt{1-\exp(-\pi/\epsilon)}$} in the limit of $\tauzero\rightarrow\infty$. For positive times, $\abs{a}$ and $\abs{b}$ display the decaying Stueckelberg oscillations.
    The phases $\varphi_a$ and $\varphi_b$ are both negative and are either symmetric for $\varphi_a$ or antisymmetric for $\varphi_b$ with respect to $\tau = 0$.
    Here, we have chosen $\tau_{0}=100.0$ and $\epsilon=3.0$.}
    \label{fig:1}
\end{figure}

In its most elementary version the Landau-Zener problem involves the \textit{linear} crossing of two energy levels denoted by $a$ and $b$ as indicated in \cref{fig:1} (a).
A time-independent coupling between them leads to the Schrödinger equations
\begin{subequations}\label{eq:group:a:b}
    \begin{align}
        \label{eq:a}
        i\dot{a} &= -\epsilon\tau a + b \\
        \label{eq:b}
        i\dot{b} &= \epsilon\tau b + a
    \end{align}
\end{subequations}
for the two probability amplitudes $ a = a(\tau) $ and $ b = b(\tau) $.
Here, the dot indicates the derivative with respect to the dimensionless time $\tau$ scaled in the coupling constant $\Omega$ between the two levels, and $\epsilon \equiv \alpha / \Omega^2 $ denotes the dimensionless chirp involving the steepness of the level crossing.

Throughout the article, we consider the initial conditions
\begin{subequations}\label{eq:group:initial:conditions}
    \begin{align}
        \label{eq:initial:condition:a}
        a(-\tau_{0}) &= 1\\
        \label{eq:initial:condition:b}
        b(-\tau_{0}) &= 0
    \end{align}
\end{subequations}
with $1\ll\tau_{0}$. 

\begin{figure}[htbp]
    \includegraphics[width=\columnwidth]{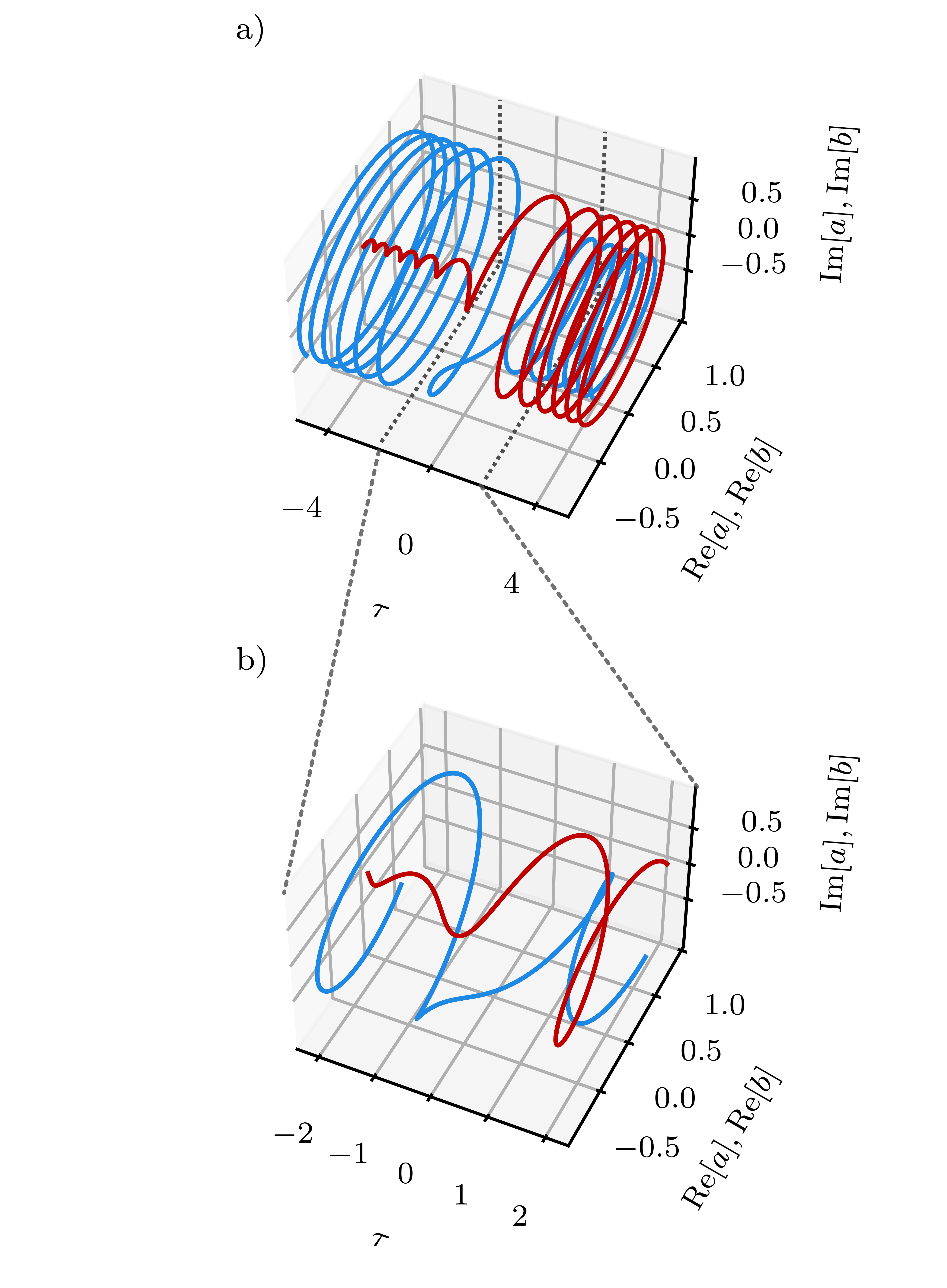}
    \centering
    \caption[]{Time evolution of the two probability amplitudes $a$ and $b$ in the complex plane. 
    For large negative times, $a$ (blue line) performs a circular motion on the unit circle. Simultaneously, $b$ (red line) starts from the origin of the complex plane and rotates around it while its amplitude increases (a).
    In the vicinity of $\tau=0$, amplified in (b), $a$ and $b$ quickly reach their asymptotic values for large positive $\tau$, while still rotating. Whereas $a$ changes the direction of rotation close to $\tau = 0$, the amplitude $b$ maintains its direction.  Here, we have chosen $\tau_{0}=100.0$ and $\epsilon=3.0$.}
    \label{fig:3d}
\end{figure}

When we solve eqs.~(\ref{eq:group:a:b}) subjected to eqs.~(\ref{eq:group:initial:conditions}) numerically, we find the time-dependence of $a$ and $b$ shown in figs. \ref{fig:1} (b) and (c), either by their absolute values and phases, or in \cref{fig:3d} as trajectories in the complex plane.

Since according to \cref{eq:group:initial:conditions}, we start from a situation where initially only the level $a$ is populated, we find in lowest approximation $a=1$ and $b=0$ till the crossing of the levels at $\tau = 0$ as depicted in \cref{fig:1} (b).
Here, the coupling is most effective and causes an abrupt reduction of $a$ to the familiar Landau-Zener value
\begin{align}
    \label{eq:LZ:value}
      \limtauzeroinfty a(\tauzero) = e^{-\frac{\pi}{2\epsilon}} \equiv a_{LZ}.
\end{align}

At the same time, the absolute value $\abs{b}$ of $b$ jumps from $b=0$ to
\begin{equation}
     \limtauzeroinfty \abs{b(\tauzero)} = \sqrt{1 - a_{LZ}^2} \equiv b_{LZ}
\end{equation}
determined by the conservation of probability
\begin{equation}
    \label{eq:conservation:prob}
    \abs{a}^2 + \abs{b}^2 = 1.
\end{equation}

Needless to say, the transition from one value to another one is not discontinuous but smooth, and is fast compared to the remaining dynamics.
Indeed, as apparent in \cref{fig:1} (b) it is concentrated around $\tau = 0$ where the non-adiabaticity of the Landau-Zener transition occurs.

In the representation of $a$ and $b$ as trajectories in the complex plane shown in \cref{fig:3d}, the two transitions from $\abs{a}=1$ to $\abs{a}=a_{LZ}$, and from $\abs{b}=0$ to $\abs{b}=b_{LZ}$ manifest themselves either in a rapid transition from a circular motion of radius unity, to one of radius $a_{LZ}$, or in a spiraling out of the origin of the complex plane to reach a circular motion of radius $b_{LZ}$.

The corresponding phases $ \varphi_a \equiv \arg\left[a\right] $ and $ \varphi_b \equiv \arg\left[b\right]$ of these motions, shown in \cref{fig:1} (c) display a dramatic difference which manifests itself most clearly in their phase velocities $\dot{\varphi}_a$ and $\dot{\varphi}_{b}$.
Indeed, $ \dot{\varphi}_a $ changes its sign in the course of the transition, corresponding to first a clockwise rotation in the complex plane for negative times, to be followed by a counter-clockwise rotation for positive times. 
In contrast, $ \dot{\varphi}_b$ is always negative leading to a clockwise rotation.
These two distinct behaviors stand out most clearly by the blue and the red line in \cref{fig:3d} (b) around $\tau=0$.

This time domain plays a special role.
In contrast to large positive and negative times where both phase velocities are large, leading to rapid oscillations around the origin, in the neighborhood of the transition both phase velocities are very small. 

Moreover from \cref{fig:1} (b), we note oscillations in $\abs{a}$ and $\abs{b}$ for positive times.
These Stueckelberg oscillations manifest themselves in the absolute values and phases of the probability amplitudes.

Whereas the probability amplitude $a$ starts from a real number, that is, from $ a(-\tau_0) = 1 $ and terminates at a real one, that is, at $ a(\tau_0) = a_{LZ} $, the asymptotic probability amplitude $ b(\tau_0) $ enjoys the non-trivial phase 
\begin{align}
    \label{eq:non:trivial:phase:phi:b}
    \varphi_b(\tau_0) \equiv \frac{\pi}{4} - \epsilon \tau_0^2 - \frac{1}{\epsilon} \ln(\sqrt{2 \epsilon}\tauzero) + \arg \Gamma\left(\frac{\ii}{2 \epsilon} \right)
\end{align}
as shown in \cref{subsection:asymptotic:values} with help of the tools developed in appendix \ref{appendix:B}.
Here, $\Gamma$ denotes the gamma function.

\section{Elementary Explanations}
\label{section:3}
In this section, we explain the features summarized in the preceding section using asymptotic solutions of the Landau-Zener equations, \cref{eq:a,eq:b}. 
In particular, we drive approximate but analytical expressions for the two probability amplitudes in the limit of large negative and large positive times. 
Moreover, we bring to light the crucial role of the logarithmic phase in triggering the transition. 

We emphasize that our asymptotic analysis, which explains \textit{qualitatively} the behavior of the two probability amplitudes, does not employ the solutions of the Landau-Zener problem in terms of parabolic cylinder functions discussed in Appendix \ref{appendix:B}. 
It is interesting, that for large negative times our approach also agrees \textit{quantitatively} with the predictions of the parabolic cylinder functions.

However, to have a quantitative agreement for large positive times the parabolic cylinder functions are necessary. 
Indeed, they determine the constants appearing in the superposition of our asymptotic solutions.

\subsection{Asymptotic solutions}

In the present section, we first motivate and then rigorously prove  solutions of the Landau-Zener equations, \cref{eq:a,eq:b}, in the limit of large negative and large positive times. 
For this purpose, we introduce elementary waves with a constant amplitude and a phase consisting of a sum of two terms which are quadratic and logarithmic in time.

\subsubsection{Elementary waves}

We get a first glimpse of the presence of a logarithmic phase in \cref{eq:group:a:b}, when we consider these equations in the limit of large times. 
For this purpose, we make the ansatz 
\begin{align}   
    \label{eq:ansatz:tau:nu}
    a(\tau) \equiv \ee^{\ii\epsilon\tau^{2}/2}\tau^{\nu}
\end{align}
which is motivated by the lowest order asymptotic expansion of the parabolic cylinder function as shown in appendix \ref{appendix:B}.

The differentiation of \cref{eq:ansatz:tau:nu} yields the relation
\begin{align}
    \label{eq:derivative:of:ansatz:tau:nu}
    \ii\dot{a} = -\epsilon\tau a + \ii\nu\frac{1}{\tau} a
\end{align}
and a comparison to \cref{eq:a} leads us to the identification
\begin{align}
    b = \ii\nu\frac{1}{\tau}a.
\end{align}

After one more differentiation, we arrive at the identity 
\begin{align}
    \ii\dot{b} = \nu\frac{1}{\tau^{2}} a + \ii\nu\frac{1}{\tau}\left(-\epsilon\tau a + \ii\nu\frac{1}{\tau}a\right)
\end{align}
where we have made use of \cref{eq:derivative:of:ansatz:tau:nu}.

A comparison to \cref{eq:b} provides us with the equation
\begin{align}
    -\nu\left(\nu - 1\right)\frac{1}{\tau^{2}} - 2\ii\epsilon\nu = 1
\end{align}
which for $\tau\rightarrow\infty$ yields
\begin{align}
    \nu = \frac{\ii}{2\epsilon}.
\end{align}

Thus, we find from \cref{eq:ansatz:tau:nu} the approximation
\begin{align}
    \label{eq:ansatz:a:ln:without:s}
    a(\tau) \cong f(\tau)
\end{align}
for the probability amplitude $a$ valid for $\tau\rightarrow\infty$. Here, we have introduced the elementary wave
\begin{subequations}\label{eq:group:ansatz:f:phase}
\begin{align}
    \label{eq:ansatz:f}
    f(\tau) \equiv \ee^{\ii\phi(\tau)},
\end{align}
which has a constant amplitude of unity.

The phase
\begin{align}
    \label{eq:phase}
    \phi(\tau) \equiv \frac{1}{2}\epsilon\tau^{2} + \frac{1}{2\epsilon}\ln\tau
\end{align}
\end{subequations}
consists of a quadratic chirp due to the linear change of the energy levels as well as a logarithmic contribution.

Due to the opposite signs of the change of the energy of the level $b$ compared to $a$, the probability amplitude $b$ is governed by the elementary wave $\fast$. 
Here, star denotes the complex conjugate and $\alpha$ and $\beta$ are constants.

Since according to \cref{eq:a}, we can find $b$ by differentiating $a$ which involves the derivative
\begin{subequations}\label{eq:group:diff:equations:f:f:ast}
    \begin{align}
    \label{eq:diff:equation:f}
    \ii\dot{f} = -\epsilon\tau f -\frac{1}{2\epsilon\tau} f
\end{align}
of $f$, $b$ must also involve $f$ in the form $1/(2\epsilon\tau) f$.

Likewise, we can use \cref{eq:b} to find $a$ by differentiating $b$ which involves $\fast$ and with the identity
\begin{align}
    \label{eq:diff:equation:f:ast}
    \ii\dot{f}^{\ast} = \epsilon\tau + \frac{1}{2\epsilon\tau}f^{\ast},
\end{align}
\end{subequations}
we recognize that $a$ must also involve $1/(2\epsilon\tau)\fast$. 

In appendix \ref{appendix:A}, we rederive the two linearly independent asymptotic solutions $u^{(+)} \cong f$ and \mbox{$u^{(-)}\cong1/(2\epsilon\tau)f^{\ast}$} from the asymptotic limit of the WKB-solution of the differential equation of second order for $a$. 

Thus, we are lead to the asymptotic expressions
\begin{subequations}\label{eq:group:sol:a:b:in:f:and:f:ast}
    \begin{align}
        \label{eq:a:approx:sol}
        a(\tau) &\cong \alpha f(\tau) + \frac{1}{2\epsilon\tau}\beta f^{\ast}(\tau)\\
        \label{eq:b:approx:sol}
        b(\tau) &\cong \beta f^{\ast}(\tau) - \frac{1}{2\epsilon\tau}\alpha f(\tau)
    \end{align}
\end{subequations}
for the two probability amplitudes $a$ and $b$ as a superpositions of the two elementary waves $f$ and $f^{\ast}$. 
Here, $\alpha$ and $\beta$ are constants.

\subsubsection{Verification by differentiation}

We now verify that \cref{eq:group:sol:a:b:in:f:and:f:ast}, satisfies \cref{eq:a,eq:b} in the limit of large times.
For this purpose, we first differentiate the expression, \cref{eq:a:approx:sol} for $a$ and use eqs.~(\ref{eq:group:diff:equations:f:f:ast}).
When we neglect terms proportional to $1/\tau^{2}$, we arrive at the relation
\begin{align}
    \label{eq:identity:a:dot:in:terms:of:f}
    \ii\dot{a} = -\epsilon\tau\alpha f - \frac{1}{2\epsilon\tau}\alpha f + \epsilon\tau\frac{1}{2\epsilon\tau}\beta f^{\ast},
\end{align}
which by subtracting and adding the term $\beta f^{\ast}/2$ takes the form
\begin{align}
    \label{eq:a:dot:f:f:ast}
    \ii\dot{a} =-\epsilon\tau\left(\alpha f+ \frac{1}{2\epsilon\tau}\beta f^{\ast}\right) + \left(\beta f^{\ast} - \frac{1}{2\epsilon\tau}\alpha f\right).
\end{align}

Finally, we recall the definitions, eqs.~ (\ref{eq:group:sol:a:b:in:f:and:f:ast}), of $a$ and $b$ to arrive indeed at \cref{eq:a}.

Likewise, we obtain by differentiating the expression, \cref{eq:b:approx:sol}, for $b$, and using the formulae, eqs.~(\ref{eq:group:diff:equations:f:f:ast}) the result
\begin{align}
    \label{eq:identity:b:dot:in:terms:of:f}
    \ii\dot{b} = \epsilon\tau\beta f^{\ast} + \frac{1}{2\epsilon\tau}\beta f^{\ast} + \epsilon\tau \frac{1}{2\epsilon\tau} \alpha f,
\end{align}
where we have neglected again terms proportional to $1/\tau^{2}$.

Next, we subtract and add the term $\alpha f/2$ and find
\begin{align}
    \ii\dot{b} = \epsilon\tau\left(\beta f^{\ast} - \frac{1}{2\epsilon\tau}\alpha f\right) + \left(\alpha f + \frac{1}{2\epsilon\tau}\beta f^{\ast}\right)
\end{align}
which with the definitions, eqs.~(\ref{eq:group:sol:a:b:in:f:and:f:ast}), of $a$ and $b$ reduces indeed to \cref{eq:b}.

\subsubsection{Summary of key ideas}

Three crucial ingredients govern  the asymptotic solution, eqs.~(\ref{eq:group:sol:a:b:in:f:and:f:ast}): (i) The two linearly independent waves $f$ and $f^{\ast}$ arising from the opposite signs of the linear chirp $\epsilon\tau$ in eqs.~(\ref{eq:group:a:b}).
(ii) A phase $\phi$ consisting of the sum of the corresponding phase $\epsilon\tau^{2}/2$ emerging from the integration of the linear chirp over time, and a logarithmic phase, and (iii) additional terms to $f$ and $f^{\ast}$ proportional to $f^{\ast}$ and $f$ with opposite signs to compensate the derivative of the logarithmic phase, \cref{eq:phase}, arising from $f$ and $f^{\ast}$.

We emphasize that there is a factor of $2$ in the denominator of the prefactor of the logarithmic phase, and consequently in these additional terms.
In the ansatz, \cref{eq:ansatz:tau:nu}, this factor of two emerges in a natural way. 
In our verification by differentiation, we had to subtract and add terms in eqs.~(\ref{eq:identity:a:dot:in:terms:of:f}) and (\ref{eq:identity:b:dot:in:terms:of:f}), either $\beta f^{\ast}/2$ or $\alpha f/2$, to recover $a$ or $b$.

\subsection{Clockwise rotation and spiral}

In the preceding section, we have derived the general asymptotic solutions, eqs.~(\ref{eq:group:sol:a:b:in:f:and:f:ast}). 
We now adjust them to the boundary conditions, eqs.~(\ref{eq:group:initial:conditions}), and therefore consider large negative times.
In this way, we obtain the asymptotic expressions for the two probability amplitudes in this time domain. 
We conclude by discussing the resulting dynamics in the complex plane.

\subsubsection{Initial conditions}

For large negative times, we employ the equations
\begin{subequations}\label{eq:group:sol:a:b:in:f:and:f:ast:new}
    \begin{align}
        \label{eq:a:approx:sol:new}
        a(-\abs{\tau}) &= \alpha_- f(-\abs{\tau}) - \frac{1}{2\epsilon\abs{\tau}}\beta_- f^{\ast}(-\abs{\tau})\\
        \label{eq:b:approx:sol:new}
        b(-\abs{\tau}) &= \beta_- f^{\ast}(-\abs{\tau}) + \frac{1}{2\epsilon\abs{\tau}}\alpha_- f(-\abs{\tau})
    \end{align}
\end{subequations}
where we have included the subscript '-' in the two constants $\alpha$ and $\beta$ in order to indicate that we are considering large \textit{negative} times. 

The initial conditions, \cref{eq:group:initial:conditions}, leads us to the two constants
\begin{subequations}\label{eq:group:alpha:minus:beta:minus}
\begin{align}
    \label{eq:alpha:minus}
    \alpha_{-} = \left[1 + \frac{1}{4\epsilon^{2}\tauzero^{2}}\right]^{-1}\frac{1}{\fminustauzero}
\end{align}
and 
\begin{align}
    \label{eq:beta:minus}
    \beta_{-} = -\left[1 + \frac{1}{4\epsilon^{2}\tauzero^{2}}\right]^{-1}\frac{1}{2\epsilon\tauzero}\frac{1}{\fastminustauzero},
\end{align}
\end{subequations}
and we obtain from eqs.~(\ref{eq:group:sol:a:b:in:f:and:f:ast:new}) together with eqs.~(\ref{eq:group:alpha:minus:beta:minus}) the expressions
\begin{subequations}\label{eq:group:approx:sol:a:b:large:neg:times}
\begin{align}
    \label{eq:approx:sol:a:large:neg:times}
    a(-\abs{\tau}) = \left(1 - \frac{1}{4\epsilon^{2}\tauzero^{2}}\right)\frac{\fabsminustau}{\fminustauzero} + \frac{1}{4\epsilon^{2}\tauzero\abs{\tau}}\frac{\fastabsminustau}{\fastminustauzero}
\end{align}
and
\begin{align}
    \label{eq:approx:sol:b:large:neg:times}
    b(-\abs{\tau}) = \frac{1}{2\epsilon\abs{\tau}}\frac{\fabsminustau}{\fminustauzero} - \frac{1}{2\epsilon\tauzero}\frac{\fastabsminustau}{\fastminustauzero}
\end{align}
\end{subequations}
which satisfy the initial conditions, \cref{eq:group:initial:conditions}. 
Here, we have neglected terms proportional to $1/(\tauzero^{2}\abs{\tau})$.

In appendix \ref{appendix:B}, we rederive \cref{eq:approx:sol:a:large:neg:times} from the parabolic cylinder functions.

\subsubsection{Motion in the complex plane}

From the definitions, eqs.~(\ref{eq:group:ansatz:f:phase}), of the wave $f$ and its phase $\phi$, we find the expression
\begin{subequations}\label{eq:group:approx:a:neg:times:with:bla}
\begin{align}
    \label{eq:approx:a:inserted}
    \frac{\fabsminustau}{\fminustauzero} = \ee^{\ii \Delta\phi(-|\tau|)}
\end{align}
with
\begin{align}
\begin{split}
    \label{eq:approx:phase:phi:a:inserted}
    \Delta \phi(-|\tau|) &\equiv \phi(-\abstau) - \phi(-\tauzero)\\ &= \frac{1}{2} \varepsilon (\tau^2 - \tau_0^2) + \frac{1}{2\varepsilon} \ln\left(\frac{|\tau|}{\tau_0}\right),
\end{split}
\end{align}
\end{subequations}
valid for large negative times.

The argument $\abs{\tau}/\tau_{0}$ of the logarithm in \cref{eq:approx:phase:phi:a:inserted} is positive, hence the logarithm is just a phase and the time evolution of $a$ given by \cref{eq:approx:a:inserted} traverses the unit circle in the complex plane with the time-dependent phase angle $\Delta\phi$, defined by \cref{eq:approx:phase:phi:a:inserted}.
This motion starts at $\tau = -\tau_{0}$ with a vanishing phase $\Delta\phi(-\tau_{0}) = 0$, that is, from the point $a(-\tauzero)=1$ on the real axis.

Since $\abs{\tau}<\tau_{0}$ both contributions in \cref{eq:approx:phase:phi:a:inserted} to $\Delta\phi(-\abs{\tau})$ are negative.
For large negative times, we can neglect the logarithm compared to the quadratic phase giving rise to the approximate phase velocity
\begin{align}
    \Delta\dot{\phi}\cong\epsilon\tau
\end{align}
which is negative for negative times, corresponding to clockwise rotation on the unit circle as shown in \cref{fig:3d} by the blue line.

According to \cref{eq:approx:sol:a:large:neg:times}, for large negative times $a$ is governed by the superposition of $\fabsminustau/\fminustauzero$ and its complex conjugate.
Hence, $a$ is determined by two contributions which rotate in opposite directions. 
Moreover, the complex conjugate is multiplied by the inverse of $1/(\tauzero\abstau)$ in order to satisfy the initial condition, \cref{eq:group:initial:conditions}. 
This prefactor is a consequence of the logarithmic phase and emerges from its derivative.

In the limit of $\tauzero\rightarrow\infty$, this contribution to $a$ disappears except at $\tau = 0$. 
For a large but finite value of $\tauzero$, this term becomes important in the neighborhood of $\tau\approx 1/\tauzero$. 
We note, that our approximation improves as $\tauzero\rightarrow\infty$.

In this asymptotic limit we find the approximate expression
\begin{align}
    \label{eq:a:negative:times:Delta}
    a(-\abstau) \cong \ee^{\ii \Delta\phi(-\abstau)}
\end{align}
where we have neglected terms of the order $1/\tauzero^{2}$ and $1/(\tauzero\abstau)$. 

Next, we address the time evolution of the probability amplitude $b$ given by \cref{eq:approx:sol:b:large:neg:times}, and consisting of a superposition of $f$ and $f^{\ast}$ required by the initial condition, \cref{eq:initial:condition:b}.
For large negative times $\tau$ but appropriately smaller than $\tau_{0}$, we can neglect the contribution proportional to $f^{\ast}$, and arrive at the approximation
\begin{align}
    b(-\abs{\tau}) \cong \frac{1}{2\epsilon\abs{\tau}}a(-\abs{\tau})
\end{align}
which is consistent with the expression derived in Ref.~\cite{Glasbrenner2025}.

Hence, $b$ is proportional to $a$ and rotates in the complex plane in the same direction as $a$, that is, in the clockwise direction. Moreover, for decreasing $\abs{\tau}$, the amplitude of $b$ increases leading us to a trajectory in the form of a spiral as shown in \cref{fig:3d}.

\begin{figure*}[htbp]
    \includegraphics[width=\textwidth]{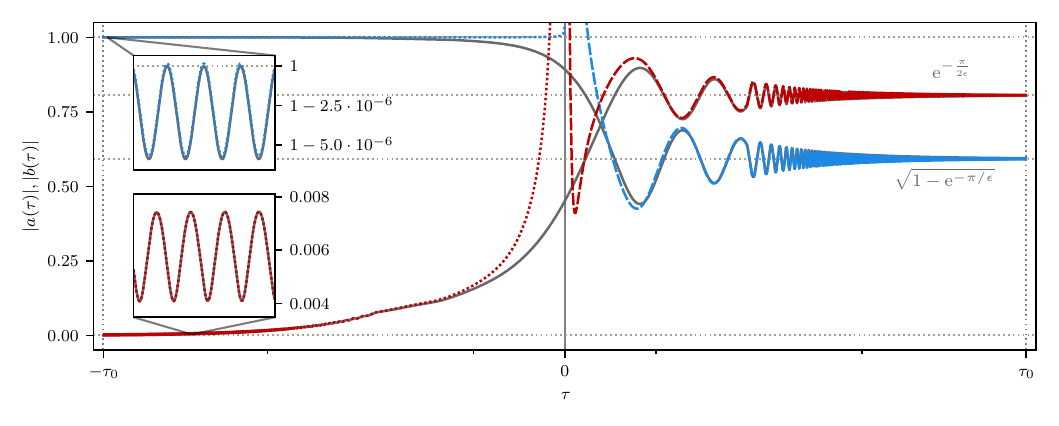}
    \centering
    \caption[]{Comparison between the exact numerical solution of eqs.~(\ref{eq:group:a:b}) for the two probability amplitudes $a$ and $b$ in their absolute values $\abs{a}$ and $\abs{b}$ and their elementary asymptotic approximations given by eqs.~(\ref{eq:group:approx:sol:a:b:large:neg:times}) and (\ref{eq:group:approx:a:neg:times:with:bla}) valid for large negative times, and \cref{eq:group:asymptotic:exp:a:b:large:times}  for large positive times.
    In order to cover the complete time domain from $-\tauzero$ to $\tauzero$ and amplify the behavior at $\tau=0$, we have chosen a logarithmic scale for $\tau$. 
    For large negative times there is an excellent agreement between the approximation and the exact result. 
    The oscillations amplified in the two insets are a consequence of the fact that $\tauzero$ is large but finite, and $a$ and $b$ are given by superpositions. 
    In the vicinity of $\tau=0$, we recognize in the approximations for both probability amplitudes $1/\tau$-singularities---they appear for negative and positive times and originate from the logarithmic phase. 
    For large positive times, the Stueckelberg oscillations result from the interference of the two contributions in \cref{eq:asymptotic:exp:a:large:times,eq:asymptotic:exp:b:large:times} and we find again an excellent agreement. 
    Since the $1/\tau$-decay appears in one of the interfering contributions the Stueckelberg oscillations decay with the same power law.
    Here, we have chosen $\tauzero=100.0$ and $\epsilon=3.0$.}
    \label{fig:3}
\end{figure*}

We conclude by comparing and contrasting the exact numerical results for the absolute values $\abs{a}$ and $\abs{b}$ to the approximations predicted by \cref{eq:approx:sol:a:large:neg:times,eq:approx:sol:b:large:neg:times} shown in \cref{fig:3}. 
We find an excellent agreement as long as $\tau$ stays appropriately away from the origin $\tau=0$.
The insets amplify the behavior for large negative times and display oscillations.
They result from the superposition of the two elementary waves $f$ and $\fast$ which is consistent with Ref.~\cite{Glasbrenner2025}.
The amplitudes of these oscillations decay for large values of $\tauzero$.

\begin{figure}[htbp]
    \includegraphics[width=\columnwidth]{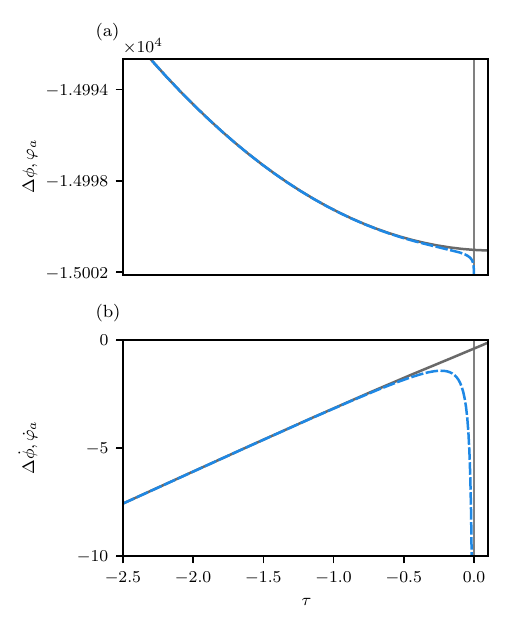}
    \centering
    \caption[]{Logarithmic phase singularity in the asymptotic expression, eqs.~(\ref{eq:group:approx:a:neg:times:with:bla}), for $a(-\abs{\tau})$. 
    In (a), we compare and contrast the approximate phase $\Delta\phi(-\abstau)$ (dashed blue) and the exact phase $\varphi_a$ (solid gray) of the probability amplitude $a$ and in (b) the corresponding phase velocities $\Delta\dot{\phi}(-\abstau)$ and $\dot{\varphi}_a$.
    Although, \cref{eq:group:approx:a:neg:times:with:bla} is valid only for \textit{large} negative times, we show $\Delta\phi(-\abstau)$ and $\Delta\dot{\phi}(-\abstau)$ in the complete domain $-\tauzero\leq\tau<0$.
    In the neighborhood of $-\tau_{0}$ the phase $\Delta\phi(-\abstau)$ is quadratic in $\tau$ giving rise to a phase velocity $\Delta\dot{\phi}(-\abstau)$ which is linear in $\tau$.
    This behavior is a consequence of the dominance of the quadratic phase compared to the logarithm.
    In this time domain, $\Delta\phi(-\abstau)$ is an excellent approximation of $\varphi_a$.
    However, close to the origin that is $\tau\lesssim 0$, the logarithm dominates giving rise to a logarithmic phase singularity and a $1/\tau$-singularity in the phase velocity.
    Here, the two functions strongly deviate since the exact phase remains finite. 
    In this figure, we have chosen $\tau_{0}=100.0$ and $\epsilon=3.0$.}
    \label{fig:2}
\end{figure}

\subsection{The transition: A heuristic argument}
\label{eq:subsection:the:transition}

It is instructive to study the behavior of the asymptotic expression, eqs.~(\ref{eq:group:approx:a:neg:times:with:bla}) for $a$ in the neighborhood of $\tau=0$ where it is not valid. 
Indeed, in this domain we can neglect the quadratic phase contributions to $\Delta\phi$ compared to the logarithm, that is 
\begin{align}
    \Delta\phi(-\abs{\tau}) \cong \frac{1}{2\epsilon}\ln\left(\frac{\abs{\tau}}{\tau_{0}}\right),
\end{align}
giving rise to a $1/\tau$-singularity in the phase velocity $\Delta\dot{\phi}$.

In \cref{fig:2}, we depict the phase $\Delta\phi(-\abs{\tau})$ given by \cref{eq:approx:phase:phi:a:inserted} as well the phase velocity $\Delta\dot{\phi}(-\abs{\tau})$.
In the neighborhood of $-\tauzero\lesssim\tau$, we note the quadratic dependence of $\Delta\phi$ and the linear increase of $\Delta\dot{\phi}$ with $\tau$.
Close to the origin, we find a logarithmic phase singularity in $\Delta \phi$ and a $1/\tau$-singularity in $\Delta\dot{\phi}$.

The curve for $\Delta\phi$ and $\Delta\dot{\phi}$ agree well with the exact curves for large negative times but deviate for $\tau\lesssim 0$.
In particular, the exact one remains finite at $\tau = 0$. 

Despite the difference between the two curves, the logarithmic phase contains the essence of the Landau-Zener transition from $a=1$ to $a_{LZ}$.
In order to bring out this fact most clearly, we note from the definition, \cref{eq:phase}, of the phase $\phi$ of the wave $f$ that the argument of the logarithm is positive for positive times. However, due to the initial conditions, \cref{eq:group:initial:conditions}, the initial time $-\tau_{0}$ enters and the argument is now negative which 
with the functional equation of the logarithm yields the relation
\begin{align}
    \label{eq:log:identity}
    \ln\left(-\frac{|\tau|}{\tau_0}\right) = \ln(-1) + \ln\left(\frac{|\tau|}{\tau_0}\right).
\end{align}

Since the logarithm of $-1$ is multi-valued, that is \mbox{$\ln(-1) = \ii(\pi+2n\pi)$}, we have to choose the physically correct branch cut. 
The initial condition $a(-\tauzero)=1$ and the conservation of probability enforces $a$ to decay, that is $\abs{a}^{2}\leq 1$ as $\tau\rightarrow\infty$.
The only choice consistent with a decaying probability amplitude is $\ln(-1)=\ii\pi$, and corresponds to analytically continuing the time $\tau$ through the upper half of the complex plane. 
The branch $\ln(-1)=-\ii\pi$, we lead us to $\abs{a}^{2}\geq 1$ as $\tau\rightarrow\infty$. 

Hence, we find, extending \cref{eq:group:approx:a:neg:times:with:bla}, to large positive times, the relation
\begin{align}
    \label{eq:connection:a:neg:pos:times}
    a(\abs{\tau}) = \ee^{-\frac{\pi}{2 \varepsilon}} a(-\abs{\tau}),
\end{align}
and thus for $\tau = \tauzero$ with the initial condition, \cref{eq:initial:condition:a}, the Landau-Zener formula, \cref{eq:LZ:value}.

Hence, in the transition from negative to positive times corresponding to a transition in the argument of the logarithm from positive real numbers to negative ones, the amplitude is reduced to the Landau-Zener probability amplitude $a_{LZ}$, given by \cref{eq:LZ:value}. The elementary approximation, \cref{eq:a:negative:times:Delta}, displays a jump from the radius $\abs{a}=1$ to $a_{LZ}$.

\subsection{Stueckelberg oscillations}

In the preceding section, we have presented a heuristic argument for the Landau-Zener formula based on the logarithmic phase singularity at $\tau=0$. 
It resulted from extending the argument of the logarithm from positive to negative numbers and choosing the branch cut consistent with the normalization condition. 
In this section, we provide a more rigorous analysis supporting this argument.

\subsubsection{The past determines the future}
The coupling between the two levels is most efficient at $ \tau = 0 $, when they cross, promoting the amplitude $ b $ to a substantial non-vanishing value.
As a result, the parameter $\beta$ in the asymptotic solution, eqs.~(\ref{eq:group:sol:a:b:in:f:and:f:ast}), is non-vanishing for large positive times, and both amplitudes $ a $ and $ b $ involve superpositions of $ f $ and $ f^\ast $ giving rise to the Stueckelberg oscillations. 

In order to derive the corresponding constants $\alpha_+$ and $\beta_+$ for the asymptotic solution, eqs.~(\ref{eq:group:sol:a:b:in:f:and:f:ast}), we recall that two linearly independent exact solutions $u=u(\tau)$ and $v=v(\tau)$ of the system, eqs.~(\ref{eq:group:a:b}), form the basis for the general expression
\begin{align}
    \label{eq:a:in:terms:u:v}
    a(\tau) = \mathcal{A} u(\tau) + \mathcal{B} v(\tau) 
\end{align}
for the probability amplitude $a$.
Here $\mathcal{A}$ and $\mathcal{B}$ are constants determined by the initial conditions, eqs.~(\ref{eq:group:initial:conditions}).

In the limit of large negative times, \cref{eq:a:in:terms:u:v}, in the form 
\begin{align}
    a(-\abstau) = \mathcal{A}u(-\abstau) + \mathcal{B}v(-\abstau)
\end{align}
must reduce to \cref{eq:a:approx:sol:new} and thus we find the identifications
\begin{subequations}\label{eq:group:u:v:tau}
\begin{align}
    \label{eq:negative:times:u}
    u(-\abs{\tau}) \cong f(-\abs{\tau})
\end{align}
and 
\begin{align}
    v(-\abs{\tau}) \cong -\frac{1}{2\epsilon\abs{\tau}}f^{\ast}(-\abs{\tau})
\end{align}
\end{subequations}
together with
\begin{align}
\label{eq:constants}
    \mathcal{A}\cong\alpha_{-}\quad\text{and}\quad\mathcal{B}\cong\beta_{-}.
\end{align}

However, for large positive times, the representation
\begin{align}
    a(\abstau) = \mathcal{A}u(\abstau) + \mathcal{B}(\abstau)
\end{align}
following from \cref{eq:a:in:terms:u:v} together with the relations, \cref{eq:constants}, for $\mathcal{A}$ and $\mathcal{B}$ leads us to the approximation
\begin{align}
    \label{eq:a:abstau:alpha:minus:u:beta:minus:v}
    a(\abstau) = \alpha_{-}u(\abstau) + \beta_{-}v(\abstau)
\end{align}
which demonstrates that the constants $\alpha_{-}$ and $\beta_{-}$ control the asymptotic behavior not only for large \textit{negative} times, but also for large \textit{positive} times.
This argument is the justification for extending the asymptotic solution obtained in the previous section from negative to positive time while keeping the initial conditions, as expressed by \cref{eq:log:identity}.

For large positive times, $u$ and $v$ are not given by a \textit{single} elementary wave $f$ or $f^{\ast}$, defined by eqs.~(\ref{eq:group:ansatz:f:phase}), but by the linear superpositions
\begin{subequations}\label{eq:group:u:v}
    \begin{align}
    u(\abstau)\cong\gamma f(\abstau) + \delta\frac{1}{2\epsilon\abstau}f^{\ast}(\abstau)
\end{align}
and 
\begin{align}
    v(\abstau) \cong \rho f(\abstau) + \sigma \frac{1}{2\epsilon\abstau}f^{\ast}(\abstau)
\end{align}
\end{subequations}
where $\gamma$, $\delta$, $\rho$ and $\sigma$ are constants.

Hence, we find from \cref{eq:a:abstau:alpha:minus:u:beta:minus:v} and \cref{eq:group:u:v} the expression
\begin{subequations}\label{eq:group:asymptotic:a:b}
\begin{align}
\begin{split}
    \label{eq:expression:for:a:001}
    a(\abstau) &\cong \alpha_{-}\left[\gamma f(\abstau) + \delta \frac{1}{2\epsilon\abstau} f^{\ast}(\abstau)\right]\\ & + \beta_{-}\left[\rho f(\abstau) + \sigma\frac{1}{2\epsilon\abstau}f^{\ast}(\abstau)\right].
\end{split}
\end{align}

Moreover, we use \cref{eq:a} to express $b$ in terms of $a$ and its derivative $\dot{a}$, which with the relations eqs.~(\ref{eq:group:diff:equations:f:f:ast}) for the derivatives of $f$ and $f^{\ast}$, as well as the ansatz, \cref{eq:expression:for:a:001}, for $a$ leads us to the formula
\begin{align}
\begin{split}
    b(\abstau) &\cong \alpha_{-}\left(\delta f^{\ast}(\abstau) - \gamma\frac{1}{2\epsilon\abstau}f(\abstau)\right)\\ &+ \beta_{-}\left(\sigma f^{\ast}(\abstau) - \rho\frac{1}{2\epsilon\abstau}f(\abstau)\right). 
\end{split}
\end{align}
\end{subequations}

In appendix \ref{appendix:B}, we rederive this structure of the asymptotic solution for large positive times starting from the exact expression in terms of the parabolic cylinder functions.
This approach also allows us to obtain explicit expressions for $\gamma, \delta, \rho$ and $\sigma$. 
However, their knowledge it is not necessary to understand the structure of the solution. 

We conclude by substituting the definitions, eqs.~(\ref{eq:group:approx:sol:a:b:large:neg:times}), of the integration constants $\alpha_{-}$ and $\beta_{-}$ into \cref{eq:expression:for:a:001} which yields the expressions
\begin{subequations}\label{eq:group:a:and:b:approx:large:positive:times}
\begin{align}
    \begin{split}
        a(\abs{\tau}) &= \gamma\left(1 - \frac{1}{4\epsilon^{2}\tauzero^{2}}\right)\frac{\fabsplustau}{\fminustauzero} + \delta\frac{1}{2\epsilon\abs{\tau}}\frac{\fastabsplustau}{\fminustauzero}\\ & - \frac{1}{2\epsilon\tauzero}\rho\frac{\fabsplustau}{\fastminustauzero} + \sigma\frac{1}{4\epsilon^{2}\abs{\tau}\tauzero}\frac{\fastabsplustau}{\fastminustauzero}
    \end{split}
\end{align}
and 
\begin{align}
    \begin{split}
        b(\abstau) &= \delta\left(1 - \frac{1}{4\epsilon^{2}\tauzero^{2}}\right) \frac{\fastabsplustau}{\fminustauzero}
        - \gamma\frac{1}{2\epsilon\abstau}\frac{\fabsplustau}{\fminustauzero}\\ &- \sigma\frac{1}{2\epsilon\tauzero} \frac{\fastabsplustau}{\fastminustauzero}+ \rho\frac{1}{4\epsilon^{2}\abstau\tauzero}\frac{\fabsplustau}{\fastminustauzero}. 
    \end{split}
\end{align}
\end{subequations}

\subsubsection{Approximate expressions}

Next, we consider, eqs.~(\ref{eq:group:a:and:b:approx:large:positive:times}) in the limit $\tauzero\rightarrow\infty$ and arrive at the approximate but analytical expressions
\begin{subequations}\label{eq:group:asymptotic:exp:a:b:large:times}
\begin{align}
    \label{eq:asymptotic:exp:a:large:times}
    a(\abs{\tau}) = \gamma\frac{f(\abs{\tau})}{f(-\tauzero)} + \frac{1}{2\epsilon\abs{\tau}}\delta\frac{f^{\ast}(\abs{\tau})}{f(-\tauzero)}
\end{align}
and
\begin{align}
    \label{eq:asymptotic:exp:b:large:times}
    b(\abs{\tau}) = \delta\frac{f^{\ast}(\abs{\tau})}{f(-\tauzero)} - \frac{1}{2\epsilon\abs{\tau}}\gamma\frac{f(\abs{\tau})}{f(-\tauzero)}.
\end{align}
\end{subequations}
for the two probability amplitudes valid for large positive times. 

Due to the decay factor $ 1/\abstau $ multiplying $ f^\ast $ in $ a $, or $ f $ in $ b $, the superpositions are only transient, and the asymptotic values of $ a $ and $b$ are solely governed by $ f $ and $ f^\ast $, respectively. 

In \cref{fig:3}, we compare the approximate analytical expression, eqs.~(\ref{eq:group:asymptotic:exp:a:b:large:times}), to the exact numerical solution of the system eqs.~(\ref{eq:group:a:b}) for positive times.
We emphasize that there is an excellent agreement. 
In particular, our elementary approach reproduces perfectly the decaying Stueckelberg oscillations originating from the superposition in \cref{eq:group:asymptotic:exp:a:b:large:times} of $f$ and $f^{\ast}$.

\begin{table}[h!]
\centering
\begin{tabular}{@{}lcccc@{}}
\toprule
\addlinespace
 & & $ -\tauzero \leftarrow \tau $ & $\quad $ & $ \tau \rightarrow \tauzero $ \\
\addlinespace
\midrule
\addlinespace
$ a $ & $\quad $ & $ \dfrac{f(-\abs{\tau})}{f(-\tauzero)} $ & $\quad $ & $ \gamma\dfrac{f(\abs{\tau})}{f(-\tauzero)} + \dfrac{1}{2\epsilon \abs{\tau}} \delta \dfrac{f^{\ast}(\abs{\tau})}{f(-\tauzero)} $ \\
\addlinespace
\midrule
\addlinespace
$ b $ & $\quad $& $ \dfrac{1}{2\epsilon\abs{\tau}}  \dfrac{f(-\abs{\tau})}{f(-\tauzero)} $ & $\quad $ & $ \delta \dfrac{f^{\ast}(\abs{\tau})}{f(-\tauzero)} - \gamma\dfrac{1}{2\epsilon\abs{\tau}} \dfrac{f(\abs{\tau})}{f(-\tauzero)} $ \\
\addlinespace
\bottomrule
\end{tabular}
\caption{Asymptotic expressions for the probability amplitudes \mbox{$a = a(\tau)$} and \mbox{$b = b(\tau)$} in terms of the elementary waves \mbox{$f(\tau) \equiv \exp\left[\ii\left(\epsilon\tau^{2}/2 + \ln(\tau)/(2\epsilon)\right)\right]$}. Here, the constants $\gamma$ and $\delta$ are derived in appendix \ref{appendix:B}.}
\label{table:1}
\end{table}

In \cref{table:1}, we summarize the asymptotic expressions for $a$ and $b$ valid for large negative and large positive times and note that the dynamics of $a$ is dictated by $f$ for negative as well as for positive times. Since $\dot{\phi}$ is linear in $\tau$, the phase velocity changes its sign. 

In contrast, $b$ is first governed by $ f $ and then by $ f^\ast $.
Thus, the corresponding phase velocity does \textit{not} change its sign in the transition from negative to positive times.
In the domain of the transition, that is around $ \tau = 0 $, both elementary waves $ f $ and $ f^\ast $ are present and the phase velocity is small.

\subsection{Asymptotic values}
\label{subsection:asymptotic:values}

Finally, we consider the symmetric long-time limit $\tau=\tauzero\rightarrow\infty$, and find from eqs.~(\ref{eq:group:asymptotic:exp:a:b:large:times}) neglecting terms proportional to $1/\tauzero$, and recalling the definition, eqs.~(\ref{eq:group:ansatz:f:phase}), of $f$ the explicit expressions
\begin{subequations}
\begin{align}
      \limtauzeroinfty a(\tauzero) = \gamma\ee^{-\frac{\pi}{2\epsilon}}
\end{align}
and 
\begin{align}
    \label{eq:lim:tauzero:prob:amplitude:b}
    \limtauzeroinfty b(\tauzero) = \ee^{\frac{\pi}{2\epsilon}}\delta\ee^{-\ii\epsilon\tauzero^{2}}\ee^{-\frac{\ii}{\epsilon}\ln\tauzero}.
\end{align}
\end{subequations}
where we have recalled \cref{eq:connection:a:neg:pos:times}.

We note that in contrast to $a$, $b$ does not converge. 
Indeed, the phase of $b$ still depends on $\tauzero$ as expressed by \cref{eq:lim:tauzero:prob:amplitude:b}.  

When we set $\gamma = 1$ we arrive indeed at the Landau-Zener result, \cref{eq:LZ:value}. 
This derivation confirms that, \cref{eq:LZ:value} is a consequence of the logarithmic phase  contained in the waves $f$ and $f^{\ast}$.

Next, we turn to $b$ and derive from the conservation of probability, \cref{eq:conservation:prob}, with $\gamma=1$ the formula
\begin{align}
    \label{eq:exact:expression:prob:amplitude:b}
    \limtauzeroinfty b(\tau_{0}) = \sqrt{1-\ee^{-\frac{\pi}{\epsilon}}}\,\ee^{\ii\varphi_b(\tau_{0})}
\end{align}
with
\begin{align}
    \label{eq:phase:of:b}
    \varphi_b(\tauzero) = \arg\delta - 2\phi(\tauzero)
\end{align}
where we have recalled the definition, \cref{eq:phase}.

When we recall from appendix \ref{appendix:B} the identification
\begin{align}
    \label{eq:phase:delta}
    \arg\delta = \frac{\pi}{4} - \frac{1}{2\epsilon}\ln\left(2\epsilon\right) + \arg\Gamma\left(\frac{\ii}{2\epsilon}\right),
\end{align}
the asymptotic phase $\varphi_b(\tauzero)$ of $b$ consists of the difference of the phase of $\delta$, which is independent of $\tauzero$, and closely related \cite{Kayanuma1997, ivakhnenko2023} to the Stokes phase, and the dynamical phase $2\phi(\tauzero)$.
Finally, we substitute \cref{eq:phase:delta} into \cref{eq:phase:of:b} and recall \cref{eq:phase}, we arrive at \cref{eq:non:trivial:phase:phi:b}.

In summary, our approach reproduces correctly the dependence of the phase of $b$ on the asymptotic parameter $\tauzero$.
Unfortunately, it is unable to find the dependence of this phase on the dimensionless interaction parameter $\epsilon$.

\section{Conclusion and Outlook}
\label{section:5}

Decouple the two differential equations of \textit{first} order for the two probability amplitudes of the Landau-Zener problem by differentiating one of them and then substituting the other one into it!
Solve the resulting differential equation of \textit{second} order in an exact way in terms of parabolic cylinder functions! 
Finally perform in the resulting expression the limit of large times!
This three-step process summarizes the text-book derivation of the  Landau-Zener formula.

In the present article, we have pointed out a short cut which not only reproduces the key features of this effect but also brings to light its deeper origin confined in a logarithmic phase.
Indeed, the linearity of the system of differential equations
ensures that the general solution is a superposition of two exact linearly independent solutions of this system. Moreover, the boundary conditions in the infinite past determine this superposition in the infinite future.
It is for this reason, that we can extend the logarithm from positive to negative arguments creating in this way a factor $\ii\pi$, and consequently the Landau-Zener factor. 
Needless to say, this insight is also contained in the asymptotic analysis of the standard approach based on parabolic cylinder functions, but only emerges after a cumbersome calculation. 

Of course, our short cut comes at a price.
We cannot completely determine the two complex-valued constants connecting the probability amplitudes in the two time domains.

However, we emphasize that the derivation of the constants was not the purpose of our article. 
Indeed, we rather wanted to bring to light the structure of the solutions of the Landau-Zener effect by solely employing elementary functions, and point out the crucial role of the logarithmic phase singularity. 

One might wonder if a similar approach would yield insight into generalizations of the Landau-Zener model discussed in this article. 
Unfortunately, this task goes beyond the scope of article.

\section*{Dedication}
We dedicate this article to our dear friend Prof. Dr. Ernst Maria Rasel — an amazing experimentalist, a deep thinker and above all, a wonderful human being — on the occasion of his 60th birthday. 
0ur interest in and our fresh analysis of the familiar Landau-Zener effect has been partially stimulated by his pathbreaking experiments on atoms in accelerated optical lattices.
We hope that he will enjoy our modest birthday present. 

Indeed, one of us (WPS) has been enormously fortunate to have been able to collaborate with him on numerous fascinating problems in atom optics and atoms in micro-gravity.
In our discussions we have learned so much from him, and his insights have always triggered new projects.
For this reason, we look forward very much to collaborating for at least another 30 years. 

\begin{acknowledgments}
We thank  D. Fabian, V. V. Kocharovsky and S. Varro for many fruitful discussions.
W.P.S. is most grateful to Texas A\&M University for a Faculty Fellowship at the Hagler Institute for Advanced Study at Texas A\& M University and to Texas A\&M AgriLife for the support of this work. 
\end{acknowledgments}

\section*{Data Availability}
The data that support the findings of this study are available within the article.

\section*{Author declarations}
\subsection*{Conflict of interest}
The authors have no conflicts to disclose.
\subsection*{Author contributions}
\textbf{Eric P. Glasbrenner:} Conceptualization (equal); Formal analysis
(equal); Investigation (lead); Methodology (equal); Validation (lead); Visualization (lead);
Writing – original draft (equal); Writing – review \& editing (equal)
\textbf{Wolfgang P. Schleich:}
Conceptualization (equal); Formal analysis
(equal); Investigation (supporting); Methodology (equal); Validation (supporting); Visualization (supporting);
Writing – original draft (equal); Writing – review \& editing (equal)

\appendix

\onecolumngrid

\section{Asymptotic solutions in terms of parabolic cylinder functions}
\label{appendix:B}

In this appendix, we derive the asymptotic expressions, eqs.~(\ref{eq:approx:sol:a:large:neg:times}) and (\ref{eq:group:asymptotic:a:b}), valid in the limit of large negative and large positive times, from the exact expression \cite{Glasbrenner2025} for $a$ in terms of the parabolic cylinder functions.
This approach is made possible by their asymptotic formulas. 
We first summarize the key equations and then consider the two distinct domains of time.

\subsection{General ingredients}

We start by recalling the exact expression \cite{Glasbrenner2025} 
\begin{align}
    \label{eq:appendix:def:a:in:terms:of:z}
    a(z) \equiv \frac{\mathcal{Z}(z)}{\mathcal{Z}_{0}}
\end{align}
for the probability amplitude $a$ where we have introduced the abbreviation
\begin{align}
    \label{eq:appendix:definition:Z}
    \mathcal{Z}(z) \equiv \parabolicfunction{\nu+1}{z_0}\parabolicfunction{\nu}{z} + \parabolicfunction{\nu+1}{-z_0}\parabolicfunction{\nu}{-z}
\end{align}
in terms of the parabolic cylinder functions $\mathcal{D}_{\nu}$ and $\mathcal{D}_{\nu+1}$ \cite{gradshteyn2014} of the order
\begin{align}
    \label{eq:appendix:group:def:nu}
    \nu \equiv -1 - \frac{\ii}{2\epsilon}\quad\text{and}\quad \nu + 1 = -\frac{\ii}{2\epsilon}
\end{align}
together with 
\begin{align}
    \mathcal{Z}_{0} \equiv \mathcal{Z}(-\zzero).
\end{align}

Moreover, 
\begin{subequations}\label{eq:appendix:group:complex:scaling}
\begin{align}  
    \label{eq:appendix:complex:scaling:plus}
    z \equiv \sqrt{2\epsilon}\abs{\tau}\ee^{\ii\pi/4}
\end{align}
and 
\begin{align}
    -z = \sqrt{2\epsilon}\abs{\tau}\ee^{-\ii 3\pi/4}
\end{align}
\end{subequations}
denote the scaled time rotated by $\pi/4$ and $-3\pi/4$ corresponding to positive and negative times.

To perform the two asymptotic limits, we employ the two formulae \cite{gradshteyn2014}
\begin{subequations}\label{eq:appendix:group:pcf:minus:5:pi:over:4:minus:pi:over:4}
\begin{align}
    \label{eq:appendix:definition:pcf:asymptotic:3:pi:over:4}
    \parabolicfunction{\nu}{z} \cong \ee^{-z^{2}/4}z^{\nu}
\end{align}
and 
\begin{align}
    \label{eq:appendix:definition:pcf:asymptotic:minus:5:pi:over:4:minus:pi:over:4}
    \parabolicfunction{\nu}{z} \cong \ee^{-z^{2}/4}z^{\nu} + \kappa_{\nu}\ee^{-\ii(\nu + 1)\pi}\ee^{z^{2}/4}z^{-(\nu + 1)}
\end{align}
\end{subequations}
valid for $\abs{\arg[z]}<3\pi/4$ and $-5\pi/4 < \arg[z]<-\pi/4$.
Here, the coefficient
\begin{align}
    \label{eq:appendix:definition:kappa:nu}
    \kappa_{\nu} \equiv \frac{\sqrt{2\pi}}{\Gamma(-\nu)}
\end{align}
contains the Gamma function $\Gamma$. 

\subsection{Large negative times}
We first focus on large negative times and consider the limit of 
\begin{align}
    \label{eq:appendix:Z:of:minus:z:pcf}
    \mathcal{Z}(-z) = \parabolicfunction{\nu+1}{z_0}\parabolicfunction{\nu}{-z} + \parabolicfunction{\nu+1}{-z_0}\parabolicfunction{\nu}{z}
\end{align}
following from \cref{eq:appendix:definition:Z} which with the asymptotic expressions, eqs.~(\ref{eq:appendix:group:pcf:minus:5:pi:over:4:minus:pi:over:4}), takes the form
\begin{align}
    \begin{split}
        \label{eq:appendix:Z:of:minus:z:all:terms}
        \mathcal{Z}(-z) = A_{-} + B_{-} + C_{-} + D_{-}
    \end{split}
\end{align}
where we have introduced the abbreviations
\begin{subequations}
    \begin{align}
        \label{eq:appendix:A:minus}
        A_{-} &\equiv \ee^{-\ii\pi\nu}\frac{1}{z}\ee^{-\left(z^{2} + z_{0}^{2}\right)/4} \left(z z_{0}\right)^{\nu + 1}, \\
        \label{eq:appendix:B:minus}
        B_{-} &\equiv \kappa_{\nu}\ee^{\left(z^{2} - z_{0}^{2}\right)/4}\left(\frac{\zzero}{z}\right)^{\nu + 1},\\
        \label{eq:appendix:C:minus}
        C_{-} &\equiv -\ee^{-\ii\pi\nu}\frac{1}{z}\ee^{-\left(z^{2} + z_{0}^{2}\right)/4} \left(z z_{0}\right)^{\nu + 1},\\
        \label{eq:appendix:D:minus}
        D_{-} &\equiv \kappa_{\nu + 1} \ee^{-\left(\zsquard - \zsquardzero \right)/4}\frac{1}{z\zzero}\left(\frac{z}{\zzero}\right)^{\nu+1}.
    \end{align}
\end{subequations}

A comparison between \cref{eq:appendix:A:minus,eq:appendix:C:minus} reveals the identity
\begin{align}
    \label{eq:appendix:A:C:identity}
    A_{-} = -C_{-}
\end{align}
and the definition, \cref{eq:appendix:definition:kappa:nu}, of $\kappa_{\nu}$ leads us with the functional equation of the Gamma function \cite{gradshteyn2014} to the relation
\begin{align}
    \label{eq:appendix:kappa:relation}
    \frac{\kappa_{\nu+1}}{\kappa_{\nu}} = \frac{\Gamma(-\nu)}{\Gamma(-(\nu + 1))} = \frac{\Gamma(-(\nu + 1) + 1)}{\Gamma(-(\nu + 1))} = - (\nu + 1).
\end{align}

Thus, \cref{eq:appendix:Z:of:minus:z:all:terms} reduces with the help of \cref{eq:appendix:A:C:identity} and \cref{eq:appendix:kappa:relation} to the expression
\begin{align}
    \mathcal{Z}(-z) = \kappa_{\nu}\left[\ee^{(\zsquard - \zsquardzero)/4}\left(\frac{z}{\zzero}\right)^{-(\nu + 1)} - \frac{\nu + 1}{z\zzero}\ee^{-\left(\zsquard - \zsquardzero\right)/4}\left(\frac{z}{\zzero}\right)^{\nu + 1}\right]
\end{align}
which implies 
\begin{align}
    \label{eq:appendix:definition:Z:z:zero:negative:times}
    \mathcal{Z}_0 = \mathcal{Z}(-\zzero) = \kappa_{\nu}\left[1-\frac{\nu + 1}{\zzero^{2}}\right].
\end{align}

With the help of the expansion \mbox{$1/(1-x) \approx 1 + x$}, the probability amplitude, \cref{eq:appendix:def:a:in:terms:of:z}, reads
\begin{align}
    \label{eq:appendix:a:minus:z}
   a(-z) \cong \left[1+\frac{\nu + 1}{\zzero^{2}}\right]\ee^{(\zsquard - \zsquardzero)/4}\left(\frac{z}{\zzero}\right)^{-(\nu + 1)} - \frac{\nu + 1}{z\zzero}\ee^{-\left(\zsquard - \zsquardzero\right)/4}\left(\frac{z}{\zzero}\right)^{\nu + 1}
\end{align}
where we have neglected terms of the order $1/z^{3}$.

The definition, \cref{eq:appendix:complex:scaling:plus}, leads us to the identities
\begin{align}
    z\zzero = 2\ii\epsilon\abs{\tau}\tauzero\quad\text{and}\quad\zzero^{2} = 2\ii\epsilon\tauzero^{2}.
\end{align}

Moreover, with the definitions, \cref{eq:appendix:group:def:nu} and \cref{eq:group:ansatz:f:phase} of $\nu$ and $f$, we find 
\begin{align}
    \label{eq:appendix:definition:of:minus:a:tau}
    a(-\abs{\tau}) = \left(1 - \frac{1}{4\epsilon^{2}\tauzero^{2}}\right)\frac{\fabsplustau}{\fplustauzero} + \frac{1}{4\epsilon^{2}\tauzero\abs{\tau}}\frac{\fastabsplustau}{\fastplustauzero},
\end{align}
which is in complete agreement with \cref{eq:approx:sol:a:large:neg:times} when we note the identity
\begin{align}
    \frac{\fabsminustau}{\fminustauzero} = \frac{\fabsplustau}{\fplustauzero}.
\end{align}

\subsection{Large positive times}
We now turn to large positive times and analyze the expression, \cref{eq:appendix:definition:Z}, which with the help of the asymptotic formulae, eqs.~(\ref{eq:appendix:group:pcf:minus:5:pi:over:4:minus:pi:over:4}), takes the form
\begin{align}
    \label{eq:appendix:definition:mathcal:Z:positive:z}
    a(z) = \frac{B_{+} + D_{+}}{\mathcal{Z}_{0}} + \frac{A_{+} + C_{+}}{\mathcal{Z}_{0}}
\end{align}
where we have introduced the abbreviations
\begin{subequations}
    \begin{align}
        \label{eq:appendix:A:plus}
        A_{+} &\equiv \left[1 - \left(-1\right)^{2\nu}\right]\frac{1}{z}\ee^{-(\zsquard + \zsquardzero)/4}\left(z\zzero\right)^{\nu + 1} \\
        \label{eq:appendix:B:plus}
        B_{+} &\equiv \kappa_{\nu}\ee^{-\ii\pi(\nu + 1)}\ee^{\left(\zsquard - \zsquardzero\right)/4}\left(\frac{\zzero}{z}\right)^{\nu+1}\\
        \label{eq:appendix:C:plus}
        C_{+} &\equiv \kappa_{\nu}\kappa_{\nu+1}\frac{1}{\zzero}\ee^{\left(\zsquard + \zsquardzero \right)/4}\left(z\zzero\right)^{-(\nu + 1)}\\
        \label{eq:appendix:D:plus}
        D_{+} &\equiv \kappa_{\nu + 1}\ee^{-\ii\pi(\nu + 1)}\frac{1}{z\zzero}\ee^{-\left(\zsquard - \zsquardzero\right)/4}\left(\frac{z}{\zzero}\right)^{\nu + 1}.
    \end{align}
\end{subequations}

We start by evaluating the expression
\begin{align}
    \frac{B_{+} + D_{+}}{\mathcalZzero}= \ee^{-\ii\pi(\nu+1)}\left\{\left[1 + \frac{\nu+1}{\zzero^{2}}\right]\ee^{\left(\zsquard - \zsquardzero\right)/4}\left(\frac{z}{\zzero}\right)^{-(\nu+1)}
    - \frac{\nu+1}{z\zzero}\ee^{-\left(\zsquard - \zsquardzero \right)/4}\left(\frac{z}{\zzero}\right)^{\nu+1}\right\}
\end{align}
where we have used the relations \cref{eq:appendix:kappa:relation} and \cref{eq:appendix:definition:Z:z:zero:negative:times}, and have neglected terms of the order $1/z^{3}$.

With the help of the identity
\begin{align}
    \label{eq:appendix:exp:nu:plus:one:equal:lz}
    \ee^{-\ii\pi(\nu+1)} = \ee^{-\frac{\pi}{2\epsilon}}
\end{align}
following from the definition, \cref{eq:appendix:group:def:nu}, of $\nu$, and recalling the expression, \cref{eq:appendix:a:minus:z}, for $a(-z)$, we find
\begin{align}
    \label{eq:appendix:b:plus:D:plus}
    \frac{B_{+} + D_{+}}{\mathcalZzero}=\ee^{-\frac{\pi}{2\epsilon}}a(-z).
\end{align}

Next, we note from \cref{eq:appendix:A:plus} and \cref{eq:appendix:C:plus} that in contrast to large negative times, the terms $A_{+}$ and $C_{+}$ do not cancel each other but lead us the expression
\begin{align}
    \label{eq:appendix:A:plus:C:plus}
    \frac{A_{+} + C_{+}}{\mathcalZzero} = K_{\nu}\ee^{-\left(\zsquard + \zsquardzero\right)/4}\frac{1}{z}\left(z\zzero\right)^{\nu+1}
    +\kappa_{\nu+1}\ee^{\left(\zsquard + \zsquardzero\right)/4}\frac{1}{\zzero}\left(z\zzero\right)^{-(\nu+1)}
\end{align}
where we have introduced the abbreviation 
\begin{align}
    K_{\nu} \equiv \frac{1}{\kappa_{\nu}}\left[1-\ee^{-\ii 2\pi\nu}\right],
\end{align}
and have neglected again terms of the order $1/z^{3}$. 

With the help of the definition, \cref{eq:appendix:group:def:nu}, of $\nu$, and \cref{eq:appendix:exp:nu:plus:one:equal:lz}, we arrive at 
\begin{align}
     K_{\nu} = \frac{\Gamma(1+\frac{\ii}{2\epsilon})}{\sqrt{2\pi}}\left[1 - \ee^{-\frac{\pi}{\epsilon}}\right].
\end{align}

When we employ the functional equation
\begin{align}
    \Gamma\left(1+\frac{\ii}{2\epsilon}\right) = \frac{\ii}{2\epsilon}\Gamma\left(\frac{\ii}{2\epsilon}\right)
\end{align}
of the Gamma function, together with the decomposition
\begin{align}
\label{eq:appendix:decomp:gamma}
   \Gamma\left(\frac{\ii}{2\epsilon}\right) = \abs{\Gamma\left(\frac{\ii}{2\epsilon}\right)}\ee^{\ii\arg\left[\Gamma\left(\frac{\ii}{2\epsilon}\right)\right]}
\end{align}
in amplitude \cite{gradshteyn2014} 
\begin{align}
    \label{eq:appendix:decomp:gamma:abs}
    \abs{\Gamma\left(\frac{\ii}{2\epsilon}\right)}&=\sqrt{\frac{\pi}{\frac{1}{2\epsilon}\sinh\left(\frac{\pi}{2\epsilon}\right)}} = 2\sqrt{\pi\epsilon}\frac{\ee^{-\frac{\pi}{4\epsilon}}}{\sqrt{1-\ee^{-\frac{\pi}{\epsilon}}}}
\end{align}
and phase, we find
\begin{align}
    \label{eq:appendix:K:nu:inserted}
    K_{\nu} = \frac{\ee^{\ii\frac{\pi}{2}}}{\sqrt{2\epsilon}}\sqrt{1-\ee^{-\frac{\pi}{\epsilon}}}\ee^{\ii\arg\left[\Gamma\left(\frac{\ii}{2\epsilon}\right)\right]}\ee^{-\frac{\pi}{4\epsilon}}.
\end{align}

Similarly, we obtain for the coefficient
\begin{align}
    \kappa_{\nu+1} = \frac{\sqrt{2\pi}}{\Gamma(-\nu - 1)} = \frac{\sqrt{2\pi}}{\Gamma\left(\frac{\ii}{2\epsilon}\right)}
\end{align}
with the help of \cref{eq:appendix:decomp:gamma} and \cref{eq:appendix:decomp:gamma:abs} the form
\begin{align}
    \label{eq:appendix:kappa:nu:plus:one:inserted}
    \kappa_{\nu+1} = \frac{1}{\sqrt{2\epsilon}}\sqrt{1-\ee^{-\frac{\pi}{\epsilon}}}\ee^{-\ii\arg\left[\Gamma\left(\frac{\ii}{2\epsilon}\right)\right]}\ee^{\frac{\pi}{4\epsilon}}.
\end{align}

Finally, we recall the definition, \cref{eq:appendix:complex:scaling:plus}, of $z$, 
and find the expressions
\begin{subequations}\label{eq:appendix:group:z:z:zero}
\begin{align}
    \frac{1}{z}\left(z\zzero\right)^{\nu+1} = \frac{\ee^{-\ii\frac{\pi}{4}}}{\sqrt{2\epsilon}\abs{\tau}} \ee^{\frac{\pi}{4\epsilon}} \left(2\epsilon\abs{\tau}\tauzero\right)^{-\frac{\ii}{2\epsilon}}
\end{align}
and 
\begin{align}
     \frac{1}{\zzero}\left(z\zzero\right)^{-(\nu+1)} = \frac{\ee^{-\ii\frac{\pi}{4}}}{\sqrt{2\epsilon}\tauzero} \ee^{-\frac{\pi}{4\epsilon}} \left(2\epsilon\abs{\tau}\tauzero\right)^{\frac{\ii}{2\epsilon}}.
\end{align}
\end{subequations}

Next, we substitute \cref{eq:appendix:complex:scaling:plus}, \cref{eq:appendix:K:nu:inserted}, \cref{eq:appendix:kappa:nu:plus:one:inserted} and \cref{eq:appendix:group:z:z:zero} into \cref{eq:appendix:A:plus:C:plus} and obtain 
\begin{align}
    \begin{split}
    \label{eq:appendix:A:plus:C:plus:inserted}
    \frac{A_{+} + C_{+}}{\mathcal{Z}(-\zzero)} &= \frac{1}{2\epsilon\abs{\tau}}\chi(\epsilon)\ee^{-\ii\epsilon\left(\tau^{2} + \tauzero^{2}\right)/2}\left(\abs{\tau}\tauzero\right)^{-\frac{\ii}{2\epsilon}}
    \\ &+\frac{1}{2\epsilon\tauzero}\chi^{\ast}(\epsilon)\ee^{\ii\epsilon\left(\tau^{2} + \tauzero^{2}\right)/2}\left(\abs{\tau}\tauzero\right)^{\frac{\ii}{2\epsilon}}
    \end{split}
\end{align}
with the abbreviation
\begin{align}
    \chi(\epsilon) \equiv \sqrt{1-\ee^{-\frac{\pi}{\epsilon}}}\ee^{\ii\pi/4}\ee^{-\frac{\ii}{2\epsilon}\ln\left(2\epsilon\right)}\ee^{\ii\arg\left[\Gamma\left(\frac{\ii}{2\epsilon}\right)\right]}.
\end{align}

We are now in the position to present the asymptotic expression for large positive times.
For this purpose, we insert \cref{eq:appendix:b:plus:D:plus} and \cref{eq:appendix:A:plus:C:plus:inserted} into \cref{eq:appendix:definition:mathcal:Z:positive:z} and arrive at
\begin{align}
    \label{eq:appendix:a:positive:times}
    a(\abstau) = \ee^{-\frac{\pi}{2\epsilon}}a(-\abs{\tau}) + \frac{1}{2\epsilon\abs{\tau}}\chi(\epsilon)\frac{\fastabsplustau}{f(\tauzero)} + \frac{1}{2\epsilon\tauzero}\chi^{\ast}(\epsilon)\frac{\fabsplustau}{\fast(\tauzero)}
\end{align}
where we have recalled the definition, \cref{eq:group:ansatz:f:phase}, of $f$. 

When we recall the expression, \cref{eq:appendix:definition:of:minus:a:tau}, for $a(-\abstau)$ this formula takes the form 
\begin{align}
    \label{eq:appendix:a:positive:times:long:version}
     a(\abstau) = \ee^{-\frac{\pi}{2\epsilon}}\left[\left(1 - \frac{1}{4\epsilon^{2}\tauzero^{2}}\right)\frac{\fabsplustau}{\fplustauzero} + \frac{1}{4\epsilon^{2}\tauzero\abs{\tau}}\frac{\fastabsplustau}{\fastplustauzero}\right] + \frac{1}{2\epsilon\abs{\tau}}\chi(\epsilon)\frac{\fastabsplustau}{f(\tauzero)} + \frac{1}{2\epsilon\tauzero}\chi^{\ast}(\epsilon)\frac{\fabsplustau}{\fast(\tauzero)}.
\end{align}

With the help of the relations
\begin{align}
    \ee^{-\frac{\pi}{2\epsilon}}\frac{\fabsplustau}{\fplustauzero} = \frac{\fabsplustau}{\fminustauzero},\quad\ee^{\frac{\pi}{2\epsilon}}\frac{\fastabsplustau}{\fplustauzero} = \frac{\fastabsplustau}{\fminustauzero}\quad\text{and}\quad \ee^{-\frac{\pi}{2\epsilon}}\frac{\fabsplustau}{\fastplustauzero} = \frac{\fabsplustau}{\fastminustauzero}
\end{align}
\cref{eq:appendix:a:positive:times:long:version} reads
\begin{align}
    \label{eq:appendix:a:positive:times:gamma:rho:sigma}
    a(\abs{\tau}) = \gamma\left(1 - \frac{1}{4\epsilon^{2}\tauzero^{2}}\right)\frac{\fabsplustau}{\fminustauzero} + \delta\frac{1}{2\epsilon\abs{\tau}}\frac{\fastabsplustau}{\fminustauzero} - \frac{1}{2\epsilon\tauzero}\left(\rho\frac{\fabsplustau}{\fastminustauzero} + \sigma\frac{1}{2\epsilon\abs{\tau}}\frac{\fastabsplustau}{\fastminustauzero}\right).
\end{align}
where 
\begin{align}
    \gamma \equiv 1\,\,\text{and}\,\,\delta\equiv \ee^{-\frac{\pi}{2\epsilon}}\chi\,\,\text{and}\,\,\rho \equiv -\ee^{\frac{\pi}{2\epsilon}}\chi^{\ast}\,\, \text{and}\,\, \sigma \equiv -1.
\end{align}

Equation (\ref{eq:appendix:a:positive:times:gamma:rho:sigma}) is identical to expression, \cref{eq:expression:for:a:001}, derived in the main body of the text. 
However, the analysis in terms of the parabolic cylinder functions has now provided us with explicit expressions for the coefficients $\gamma, \rho, \sigma$ and $\delta$. 

\subsection{The asymptotic limit of products rather than the product of the limits} 
In the main body of this article, we have expressed the two probability amplitude $a$ and $b$ in terms of the elementary waves $f$ and $\fast$.
In this appendix, we have made the connections to the asymptotic expressions starting from the parabolic cylinder functions. 
We conclude by drawing attention to a peculiarity of this analysis. 

Indeed, a comparison between the expression, \cref{eq:appendix:def:a:in:terms:of:z}, and the decomposition, \cref{eq:a:in:terms:u:v}, allows us to make the identifications 
\begin{align}
    \label{eq:appendix:identification:u:v}
    u(z) \equiv \parabolicfunction{\nu}{z}\quad\text{and}\quad v(z) \equiv \parabolicfunction{\nu}{-z}
\end{align}
with 
\begin{align}
    \label{eq:appendix:matcal:A:B}
    \mathcal{A} \equiv \frac{\parabolicfunction{\nu + 1}{\zzero}}{\mathcal{Z}_{0}}\quad\text{and}\quad\mathcal{B}\equiv\frac{\parabolicfunction{\nu + 1}{-\zzero}}{\mathcal{Z}_{0}}.
\end{align}

However, we emphasize that the asymptotic analysis of the proceeding sections shows clearly, that one cannot perform the limits in $\mathcal{A}$ and $u$ as well as in $\mathcal{B}$ and $v$ independently. 
Only the sum of the product $\mathcal{A}u$ and $\mathcal{B}v$ leads to the expressions \cref{eq:appendix:definition:of:minus:a:tau} and \cref{eq:appendix:a:positive:times:long:version}.
\\
\twocolumngrid

\section{Elementary waves from WKB}
\label{appendix:A}

In this article, we have brought to light the inner workings of the Landau-Zener effect by deriving asymptotic solutions of the coupled differential equations of \textit{first} order for the two probability amplitudes $a$ and $b$. 
These expressions are linear combinations of $f$ and $f^{\ast}/\tau$.

In this appendix, we rederive these solutions from the WKB-waves approximating the linear differential equation of \textit{second} order
\begin{align}
    \label{eq:second:order:ode}
    \ddot{a} + \left[ \left(\epsilon\tau\right)^{2} + 1 -\ii\epsilon\right]a = 0
\end{align}
for the probability amplitude $a$ following by differentiation of \cref{eq:a} and using \cref{eq:b}.

For this purpose, we start from the elementary WKB solution \cite{schleich2015}
\begin{align}
    u^{(\pm)}(\tau) \equiv \frac{\mathcal{N}^{(\pm)}}{\sqrt{\lambda(\tau)}}\ee^{\pm\ii\theta(\tau)}
\end{align}
in terms of the analogues of the momentum
\begin{align}
    \label{eq:appendixwkb:lambda}
    \lambda(\tau) \equiv \sqrt{\left(\epsilon\tau\right)^{2} + 1 - \ii\epsilon}
\end{align}
and the action
\begin{align}
    \label{eq:appendix:B:theta}
    \theta(\tau) \equiv \int^{\tau}\dd\tau^{\prime}\lambda(\tau^{\prime}).
\end{align}
Here,  $\mathcal{N}^{(\pm)}$ are constants to be determined later. 

The limit of large times allows us to expand the square root, \cref{eq:appendixwkb:lambda}, defining $\lambda$, that is
\begin{align}
    \lambda(\tau) \cong \epsilon\tau + \frac{1}{2\epsilon\tau} - \frac{\ii}{2\tau}
\end{align}
and substitute it into the definition, \cref{eq:appendix:B:theta}, of $\theta$ which after integration leads us to the expression
\begin{align}
    \theta(\tau) \cong \frac{1}{2}\epsilon\tau^{2} + \frac{1}{2\epsilon}\ln\left(\tau\right) - \frac{\ii}{2}\ln\left(\tau\right).
\end{align}

For the time-dependent amplitude of $u^{(\pm)}$, it suffices to retain the lowest order, that is
\begin{align}
    \frac{1}{\sqrt{\lambda(\tau)}}\cong \frac{1}{\sqrt{\epsilon\tau}},
\end{align}
leading us to the approximations 
\begin{align}
    u^{(+)}(\tau) \cong \frac{\mathcal{N}^{(+)}}{\sqrt{\epsilon}}\ee^{\ii\epsilon\tau^{2}/2}\ee^{\frac{\ii}{2\epsilon}\ln\left(\tau\right)}
\end{align}
and
\begin{align}
    u^{(-)}(\tau) \cong \frac{\mathcal{N}^{(-)}}{\sqrt{\epsilon}\tau}\ee^{-\ii\epsilon\tau^{2}/2}\ee^{-\frac{\ii}{2\epsilon}\ln\left(\tau\right)}.
\end{align}

For the choice of $\mathcal{N}^{(+)}\equiv\sqrt{\epsilon}$ we find 
\begin{align}
    u^{(+)} \cong f,
\end{align}
and for $\mathcal{N}^{(-)}\equiv1/(2\sqrt{\epsilon})$
\begin{align}
    u^{(-)} \cong \frac{1}{2\epsilon\tau}f^{\ast}
\end{align}
Here, we have recalled the definitions, eqs.~(\ref{eq:group:ansatz:f:phase}) of $f$.   

\newpage

\nocite{*}

\bibliography{references}

@Article{Landau1932b,
author = {Landau, L. D.},
title = {{A} theory of energy transfer. {II}},
editor = {D. T. Haar},
journal = {Collected Papers of L.D. Landau},
publisher = {Pergamon},
pages = {63-66},
year = {1965},
isbn = {978-0-08-010586-4},
doi = {https://doi.org/10.1016/B978-0-08-010586-4.50014-6},
url = {https://www.sciencedirect.com/science/article/pii/B9780080105864500146},
}

@Article{Landau1932a,
author = {Landau, L. D.},
title = {Zur {T}heorie der {E}nergieübertragung {II}},
journal = {Sov. Phys.},
year = {1932},
volume = {2},
pages = {46-51},
number = {833},
doi = {10.1098/rspa.1932.0165},
url = {https://royalsocietypublishing.org/doi/abs/10.1098/rspa.1932.0165},
}

@Article{Zener1932,
author = {Zener, Clarence},
title = {Non-adiabatic crossing of energy levels},
journal = {Proc. R. Soc. London A},
year = {1932},
volume = {137},
pages = {696-702},
number = {833},
doi = {10.1098/rspa.1932.0165},
url = {https://royalsocietypublishing.org/doi/abs/10.1098/rspa.1932.0165}
}

@article{Majorana1932,
author={Majorana, Ettore},
title={Atomi orientati in campo magnetico variabile},
journal={Il Nuovo Cimento},
year={1932},
volume={9},
number={2},
pages={43-50},
issn={1827-6121},
doi={10.1007/BF02960953},
url={https://doi.org/10.1007/BF02960953}
}

@article{Stueckelberg1932,
author={Stueckelberg, E. C. G.},
title={Theorie der unelastischen {S}t{\"o}sse zwischen {A}tomen},
journal={Helvetica Physica Acta},
year={1932},
publisher={Birkh{\"a}user},
volume={5},
number={VI},
pages={369},
issn={0018-0238},
doi={10.5169/seals-110177},
url={https://doi.org/10.5169/seals-110177}
}

@article{Hill1953,
  title = {Nuclear {C}onstitution and the {I}nterpretation of {F}ission {P}henomena},
  author = {Hill, David Lawrence and Wheeler, John Archibald},
  journal = {Phys. Rev.},
  volume = {89},
  issue = {5},
  pages = {1102--1145},
  numpages = {0},
  year = {1953},
  publisher = {American Physical Society},
  doi = {10.1103/PhysRev.89.1102},
  url = {https://link.aps.org/doi/10.1103/PhysRev.89.1102}
}

@article{Dykhne1960,
  author    = {A. M. Dykhne},
  title     = {Adiabatic {P}erturbation of {D}iscrete {S}pectrum {S}tates},
  journal   = {Sov. Phys. JETP},
  volume    = {11},
  pages     = {411--415},
  year      = {1960},
  note      = {Zh. Eksp. Teor. Fiz. \textbf{38}, 570 (1960)}
}

@article{Dykhne1962,
  author    = {A. M. Dykhne},
  title     = {Transition {P}robability for {N}onstationary {P}roblems},
  journal   = {Sov. Phys. JETP},
  volume    = {14},
  pages     = {941--943},
  year      = {1962},
  note      = {Zh. Eksp. Teor. Fiz. \textbf{41}, 1324 (1961)}
}

@article{Davis1976,
    author = {Davis, Jon P. and Pechukas, Philip},
    title = {Nonadiabatic transitions induced by a time‐dependent {H}amiltonian in the semiclassical/adiabatic limit: {T}he two‐state case},
    journal = {The Journal of Chemical Physics},
    volume = {64},
    number = {8},
    pages = {3129-3137},
    year = {1976},
    month = {04},
    issn = {0021-9606},
    doi = {10.1063/1.432648},
    url = {https://doi.org/10.1063/1.432648}
}

@article{Zhu1992,
    author = {Zhu, Chaoyuan and Nakamura, Hiroki},
    title = {Stokes constants for a certain class of second‐order ordinary differential equations},
    journal = {Journal of Mathematical Physics},
    volume = {33},
    number = {8},
    pages = {2697-2717},
    year = {1992},
    issn = {0022-2488},
    doi = {10.1063/1.529591},
    url = {https://doi.org/10.1063/1.529591}
}

@article{Berry1993,
doi = {10.1088/0305-4470/26/18/034},
url = {https://dx.doi.org/10.1088/0305-4470/26/18/034},
year = {1993},
publisher = {},
volume = {26},
number = {18},
pages = {4737},
author = {M V Berry and R Lim},
title = {Universal transition prefactors derived by superadiabatic renormalization},
journal = {Journal of Physics A: Mathematical and General}
}

@article{Vitanov1996,
  title = {Landau-{Z}ener model: {E}ffects of finite coupling duration},
  author = {Vitanov, N. V. and Garraway, B. M.},
  journal = {Phys. Rev. A},
  volume = {53},
  issue = {6},
  pages = {4288--4304},
  numpages = {0},
  year = {1996},
  publisher = {American Physical Society},
  doi = {10.1103/PhysRevA.53.4288},
  url = {https://link.aps.org/doi/10.1103/PhysRevA.53.4288}
}

@article{Kayanuma1997,
  title = {Stokes phase and geometrical phase in a driven two-level system},
  author = {Kayanuma, Yosuke},
  journal = {Phys. Rev. A},
  volume = {55},
  issue = {4},
  pages = {R2495--R2498},
  numpages = {0},
  year = {1997},
  month = {Apr},
  publisher = {American Physical Society},
  doi = {10.1103/PhysRevA.55.R2495},
  url = {https://link.aps.org/doi/10.1103/PhysRevA.55.R2495}
}

@article{Teranishi1997,
    author = {Teranishi, Yoshiaki and Nakamura, Hiroki},
    title = {Semiclassical theory of time-dependent curve crossing problems},
    journal = {The Journal of Chemical Physics},
    volume = {107},
    number = {6},
    pages = {1904-1914},
    year = {1997},
    month = {08},
    issn = {0021-9606},
    doi = {10.1063/1.474541},
    url = {https://doi.org/10.1063/1.474541}
}

@article{Peik1997,
  title = {Bloch oscillations of atoms, adiabatic rapid passage, and monokinetic atomic beams},
  author = {Peik, Ekkehard and Ben Dahan, Maxime and Bouchoule, Isabelle and Castin, Yvan and Salomon, Christophe},
  journal = {Phys. Rev. A},
  volume = {55},
  issue = {4},
  pages = {2989--3001},
  numpages = {0},
  year = {1997},
  month = {Apr},
  publisher = {American Physical Society},
  doi = {10.1103/PhysRevA.55.2989},
  url = {https://link.aps.org/doi/10.1103/PhysRevA.55.2989}
}

@article{Vitanov1999,
  title = {Transition times in the {L}andau-{Z}ener model},
  author = {Vitanov, N. V.},
  journal = {Phys. Rev. A},
  volume = {59},
  issue = {2},
  pages = {988--994},
  numpages = {0},
  year = {1999},
  month = {Feb},
  publisher = {American Physical Society},
  doi = {10.1103/PhysRevA.59.988},
  url = {https://link.aps.org/doi/10.1103/PhysRevA.59.988}
}

@article{Wubs2005,
doi = {10.1088/1367-2630/7/1/218},
url = {https://dx.doi.org/10.1088/1367-2630/7/1/218},
year = {2005},
month = {oct},
publisher = {},
volume = {7},
number = {1},
pages = {218},
author = {Martijn Wubs and Keiji Saito and Sigmund Kohler and Yosuke Kayanuma and Peter Hänggi},
title = {Landau–{Z}ener transitions in qubits controlled by electromagnetic fields},
journal = {New Journal of Physics}
}

@article{Wubs2006,
  title = {Gauging a quantum heat bath with dissipative {L}andau-{Z}ener transitions},
  author = {Wubs, Martijn and Saito, Keiji and Kohler, Sigmund and H\"anggi, Peter and Kayanuma, Yosuke},
  journal = {Phys. Rev. Lett.},
  volume = {97},
  issue = {20},
  pages = {200404},
  numpages = {4},
  year = {2006},
  publisher = {American Physical Society},
  doi = {10.1103/PhysRevLett.97.200404},
  url = {https://link.aps.org/doi/10.1103/PhysRevLett.97.200404}
}

@inbook{Betz2006,
author={Betz, Volker
and Teufel, Stefan},
editor={Asch, Joachim
and Joye, Alain},
title={Landau-{Z}ener {F}ormulae from {A}diabatic {T}ransition {H}istories},
bookTitle={Mathematical Physics of Quantum Mechanics: Selected and Refereed Lectures from QMath9},
year={2006},
publisher={Springer Berlin Heidelberg},
address={Berlin, Heidelberg},
pages={19-32},
isbn={978-3-540-34273-1},
doi={10.1007/3-540-34273-7_4},
url={https://doi.org/10.1007/3-540-34273-7_4}
}

@article{Saito2007,
  title = {Dissipative {L}andau-{Z}ener transitions of a qubit: {B}ath-specific and universal behavior},
  author = {Saito, Keiji and Wubs, Martijn and Kohler, Sigmund and Kayanuma, Yosuke and H\"anggi, Peter},
  journal = {Phys. Rev. B},
  volume = {75},
  issue = {21},
  pages = {214308},
  numpages = {14},
  year = {2007},
  month = {Jun},
  publisher = {American Physical Society},
  doi = {10.1103/PhysRevB.75.214308},
  url = {https://link.aps.org/doi/10.1103/PhysRevB.75.214308}
}

@inbook{DeGrandi2010,
author={De Grandi, C.
and Polkovnikov, A.},
editor={Chandra, Anjan Kumar
and Das, Arnab
and Chakrabarti, Bikas K.},
title={Adiabatic {P}erturbation {T}heory: {F}rom {L}andau--{Z}ener {P}roblem to {Q}uenching {T}hrough a {Q}uantum {C}ritical {P}oint},
bookTitle={Quantum Quenching, Annealing and Computation},
year={2010},
publisher={Springer Berlin Heidelberg},
address={Berlin, Heidelberg},
pages={75-114},
isbn={978-3-642-11470-0},
doi={10.1007/978-3-642-11470-0_4},
url={https://doi.org/10.1007/978-3-642-11470-0_4}
}

@article{SHEVCHENKO2010,
title = {Landau–{Z}ener–{S}tückelberg interferometry},
journal = {Physics Reports},
volume = {492},
number = {1},
pages = {1-30},
year = {2010},
issn = {0370-1573},
doi = {https://doi.org/10.1016/j.physrep.2010.03.002},
url = {https://www.sciencedirect.com/science/article/pii/S0370157310000815},
author = {S.N. Shevchenko and S. Ashhab and Franco Nori}
}

@book{nakamura2012,
  title={Nonadiabatic {T}ransition: {C}oncepts, {B}asic {T}heories and Applications},
  author={Nakamura, H.},
  isbn={9789814329781},
  url={https://books.google.de/books?id=uj26CgAAQBAJ},
  year={2012},
  publisher={World Scientific}
}

@book{gradshteyn2014,
  title={Table of {I}ntegrals, {S}eries, and {P}roducts},
  author={Gradshteyn, I.S. and Ryzhik, I.M.},
  isbn={9781483265643},
  url={https://books.google.de/books?id=F7jiBQAAQBAJ},
  year={2014},
  publisher={Academic Press}
}

@article{Vitanov2017,
  title = {Stimulated {R}aman adiabatic passage in physics, chemistry, and beyond},
  author = {Vitanov, Nikolay V. and Rangelov, Andon A. and Shore, Bruce W. and Bergmann, Klaas},
  journal = {Rev. Mod. Phys.},
  volume = {89},
  issue = {1},
  pages = {015006},
  numpages = {66},
  year = {2017},
  month = {Mar},
  publisher = {American Physical Society},
  doi = {10.1103/RevModPhys.89.015006},
  url = {https://link.aps.org/doi/10.1103/RevModPhys.89.015006}
}

@article{Pagel2020,
  title = {Symmetric {B}loch oscillations of matter waves},
  author = {Pagel, Zachary and Zhong, Weicheng and Parker, Richard H. and Olund, Christopher T. and Yao, Norman Y. and M\"uller, Holger},
  journal = {Phys. Rev. A},
  volume = {102},
  issue = {5},
  pages = {053312},
  numpages = {15},
  year = {2020},
  month = {Nov},
  publisher = {American Physical Society},
  doi = {10.1103/PhysRevA.102.053312},
  url = {https://link.aps.org/doi/10.1103/PhysRevA.102.053312}
}

@article{Gebbe2021,
 author={Gebbe, Martina
and Siem{\ss}, Jan-Niclas
and Gersemann, Matthias
and M{\"u}ntinga, Hauke
and Herrmann, Sven
and L{\"a}mmerzahl, Claus
and Ahlers, Holger
and Gaaloul, Naceur
and Schubert, Christian
and Hammerer, Klemens
and Abend, Sven
and Rasel, Ernst M.},
title={Twin-lattice atom interferometry},
journal={Nature Communications},
year={2021},
month={May},
day={05},
volume={12},
number={1},
pages={2544},
issn={2041-1723},
doi={10.1038/s41467-021-22823-8},
url={https://doi.org/10.1038/s41467-021-22823-8}
}

@article{Glasbrenner2023,
doi = {10.1088/1361-6455/acc774},
url = {https://dx.doi.org/10.1088/1361-6455/acc774},
year = {2023},
publisher = {IOP Publishing},
volume = {56},
number = {10},
pages = {104001},
author = {Eric P. Glasbrenner and Wolfgang P. Schleich},
title = {The {L}andau–{Z}ener formula made simple},
journal = {Journal of Physics B: Atomic, Molecular and Optical Physics}
}

@article{ivakhnenko2023,
title = {Nonadiabatic {L}andau–{Z}ener–{S}tückelberg–{M}ajorana transitions, dynamics, and interference},
journal = {Physics Reports},
volume = {995},
pages = {1-89},
year = {2023},
issn = {0370-1573},
doi = {https://doi.org/10.1016/j.physrep.2022.10.002},
url = {https://www.sciencedirect.com/science/article/pii/S0370157322003702},
author = {Oleh V. Ivakhnenko and Sergey N. Shevchenko and Franco Nori}
}

@article{Kofman2023,
author={Kofman, P. O.
and Ivakhnenko, O. V.
and Shevchenko, S. N.
and Nori, Franco},
title={Majorana's approach to nonadiabatic transitions validates the adiabatic-impulse approximation},
journal={Scientific Reports},
year={2023},
volume={13},
number={1},
pages={5053},
issn={2045-2322},
doi={10.1038/s41598-023-31084-y},
url={https://doi.org/10.1038/s41598-023-31084-y}
}

@article{kofman2024,
  title = {Tuning the initial phase to control the final state of a driven qubit},
  author = {Kofman, Polina O. and Shevchenko, Sergey N. and Nori, Franco},
  journal = {Phys. Rev. A},
  volume = {109},
  issue = {2},
  pages = {022409},
  numpages = {11},
  year = {2024},
  publisher = {American Physical Society},
  doi = {10.1103/PhysRevA.109.022409},
  url = {https://link.aps.org/doi/10.1103/PhysRevA.109.022409}
}

@article{Konrad2024,
doi = {10.1088/2058-9565/ad8ae4},
url = {https://dx.doi.org/10.1088/2058-9565/ad8ae4},
year = {2024},
publisher = {IOP Publishing},
volume = {10},
number = {1},
pages = {015025},
author = {Konrad, Bernd and Efremov, Maxim},
title = {Angular {B}loch oscillations and their applications},
journal = {Quantum Science and Technology}
}

@incollection{Glasbrenner2024book,
  author       = {Eric P. Glasbrenner and Yannik Gerdes and Sándor Varró and Wolfgang P. Schleich},
  title        = {A different perspective on the {L}andau-{Z}ener dynamics},
  booktitle    = {Proceedings of the Second International Workshop on Quantum Nonstationary Systems},
  editor       = {A. Dodonov and C. C. H. Ribeiro},
  publisher    = {LF Editorial},
  year         = {2024},
  pages        = {273--297},
  edition      = {1},
  chapter      = {17},
}

@article{Fitzek2024,
  title = {Accurate and efficient {B}loch-oscillation-enhanced atom interferometry},
  author = {Fitzek, F. and Kirsten-Siem\ss{}, J.-N. and Rasel, E. M. and Gaaloul, N. and Hammerer, K.},
  journal = {Phys. Rev. Res.},
  volume = {6},
  issue = {3},
  pages = {L032028},
  numpages = {6},
  year = {2024},
  month = {Aug},
  publisher = {American Physical Society},
  doi = {10.1103/PhysRevResearch.6.L032028},
  url = {https://link.aps.org/doi/10.1103/PhysRevResearch.6.L032028}
}

@article{Ullinger2022,
    author = {Ullinger, Freyja and Zimmermann, Matthias and Schleich, Wolfgang P.},
    title = {The logarithmic phase singularity in the inverted harmonic oscillator},
    journal = {AVS Quantum Science},
    volume = {4},
    number = {2},
    pages = {024402},
    year = {2022},
    month = {04},
    issn = {2639-0213},
    doi = {10.1116/5.0074429},
    url = {https://doi.org/10.1116/5.0074429}
}

@article{Matsuda2025,
  title = {Exact {WKB} analysis for dynamical and geometric exponents in generalized and nonlinear {L}andau-{Z}ener transitions},
  author = {Matsuda, Tomohiro},
  journal = {Phys. Rev. A},
  volume = {112},
  issue = {3},
  pages = {032224},
  numpages = {14},
  year = {2025},
  publisher = {American Physical Society},
  doi = {10.1103/8c6k-bbpb},
  url = {https://link.aps.org/doi/10.1103/8c6k-bbpb}
}

@article{Glasbrenner2025,
  title = {Riccati equation perspective on {L}andau-{Z}ener transitions},
  author = {Glasbrenner, Eric P. and Gerdes, Yannik and Varr\'o, S\'andor and Schleich, Wolfgang P.},
  journal = {Phys. Rev. Res.},
  volume = {7},
  issue = {4},
  pages = {043208},
  numpages = {21},
  year = {2025},
  publisher = {American Physical Society},
  doi = {10.1103/nqgc-w1ns},
  url = {https://link.aps.org/doi/10.1103/nqgc-w1ns}
}

@article{bjoerkman2025,
  title = {Observation of the {T}wo-{P}hoton {L}andau-{Z}ener-{S}t\"uckelberg-{M}ajorana {E}ffect},
  author = {Bj\"orkman, Isak and Kuzmanovi\ifmmode \acute{c}\else \'{c}\fi{}, Marko and Paraoanu, Gheorghe Sorin},
  journal = {Phys. Rev. Lett.},
  volume = {134},
  issue = {6},
  pages = {060602},
  numpages = {6},
  year = {2025},
  month = {Feb},
  publisher = {American Physical Society},
  doi = {10.1103/PhysRevLett.134.060602},
  url = {https://link.aps.org/doi/10.1103/PhysRevLett.134.060602}
}

@article{Sun2025,
doi = {10.1088/1751-8121/ae0219},
url = {https://dx.doi.org/10.1088/1751-8121/ae0219},
year = {2025},
publisher = {IOP Publishing},
volume = {58},
number = {37},
pages = {37LT01},
author = {Sun, Chen},
title = {Derivation of the {L}andau–{Z}ener formula via functional equations},
journal = {Journal of Physics A: Mathematical and Theoretical}
}

@article{Berry1972,
doi = {10.1088/0034-4885/35/1/306},
url = {https://doi.org/10.1088/0034-4885/35/1/306},
year = {1972},
publisher = {},
volume = {35},
number = {1},
pages = {315},
author = {M V Berry and K E Mount},
title = {Semiclassical approximations in wave mechanics},
journal = {Reports on Progress in Physics}
}

@article{Wittig2005,
  author  = {Wittig, Curt},
  title   = {The {L}andau--{Z}ener {F}ormula},
  journal = {The Journal of Physical Chemistry B},
  volume  = {109},
  number  = {17},
  pages   = {8428--8430},
  year    = {2005},
  doi     = {10.1021/jp040627u}
}

@article{Berry1990,
    author = {Berry, Michael Victor},
    title = {Histories of adiabatic quantum transitions},
    journal = {Proceedings of the Royal Society of London. A. Mathematical and Physical Sciences},
    volume = {429},
    number = {1876},
    pages = {61-72},
    year = {1990},
    issn = {0080-4630},
    doi = {10.1098/rspa.1990.0051},
    url = {https://doi.org/10.1098/rspa.1990.0051}
}

@article{Vutha2010,
doi = {10.1088/0143-0807/31/2/016},
url = {https://doi.org/10.1088/0143-0807/31/2/016},
year = {2010},
month = {feb},
publisher = {},
volume = {31},
number = {2},
pages = {389},
author = {Vutha, Amar C},
title = {A simple approach to the {L}andau–{Z}ener formula},
journal = {European Journal of Physics}
}

@article{Siemss2020,
  title = {Analytic theory for {B}ragg atom interferometry based on the adiabatic theorem},
  author = {Siem\ss{}, Jan-Niclas and Fitzek, Florian and Abend, Sven and Rasel, Ernst M. and Gaaloul, Naceur and Hammerer, Klemens},
  journal = {Phys. Rev. A},
  volume = {102},
  issue = {3},
  pages = {033709},
  numpages = {26},
  year = {2020},
  month = {Sep},
  publisher = {American Physical Society},
  doi = {10.1103/PhysRevA.102.033709},
  url = {https://link.aps.org/doi/10.1103/PhysRevA.102.033709}
}

@article{Oliver2005,
author = {William D. Oliver  and Yang Yu  and Janice C. Lee  and Karl K. Berggren  and Leonid S. Levitov  and Terry P. Orlando },
title = {Mach-{Z}ehnder {I}nterferometry in a {S}trongly {D}riven {S}uperconducting {Q}ubit},
journal = {Science},
volume = {310},
number = {5754},
pages = {1653-1657},
year = {2005},
doi = {10.1126/science.1119678},
URL = {https://www.science.org/doi/abs/10.1126/science.1119678},
eprint = {https://www.science.org/doi/pdf/10.1126/science.1119678}}

@article{Morsch2006,
  title = {Dynamics of {B}ose-{E}instein condensates in optical lattices},
  author = {Morsch, Oliver and Oberthaler, Markus},
  journal = {Rev. Mod. Phys.},
  volume = {78},
  issue = {1},
  pages = {179--215},
  numpages = {0},
  year = {2006},
  month = {Feb},
  publisher = {American Physical Society},
  doi = {10.1103/RevModPhys.78.179},
  url = {https://link.aps.org/doi/10.1103/RevModPhys.78.179}
}

@article{Dziarmaga2005,
  title = {Dynamics of a {Q}uantum {P}hase {T}ransition: {E}xact {S}olution of the {Q}uantum {I}sing {M}odel},
  author = {Dziarmaga, Jacek},
  journal = {Phys. Rev. Lett.},
  volume = {95},
  issue = {24},
  pages = {245701},
  numpages = {4},
  year = {2005},
  month = {Dec},
  publisher = {American Physical Society},
  doi = {10.1103/PhysRevLett.95.245701},
  url = {https://link.aps.org/doi/10.1103/PhysRevLett.95.245701}
}

@article{Lim2025,
  title = {Partial {L}andau-{Z}ener transitions and applications to qubit shuttling},
  author = {Lima, Jonas R. F. and Burkard, Guido},
  journal = {Phys. Rev. B},
  volume = {111},
  issue = {23},
  pages = {235439},
  numpages = {7},
  year = {2025},
  month = {Jun},
  publisher = {American Physical Society},
  doi = {10.1103/y4xf-zjjx},
  url = {https://link.aps.org/doi/10.1103/y4xf-zjjx}
}

@article{Ribeiro2009,
  title = {Nuclear {S}tate {P}reparation via {L}andau-{Z}ener-{S}t\"uckelberg {T}ransitions in {D}ouble {Q}uantum {D}ots},
  author = {Ribeiro, Hugo and Burkard, Guido},
  journal = {Phys. Rev. Lett.},
  volume = {102},
  issue = {21},
  pages = {216802},
  numpages = {4},
  year = {2009},
  month = {May},
  publisher = {American Physical Society},
  doi = {10.1103/PhysRevLett.102.216802},
  url = {https://link.aps.org/doi/10.1103/PhysRevLett.102.216802}
}

@book{bender1999advanced,
  title={Advanced Mathematical Methods for Scientists and Engineers I: Asymptotic Methods and Perturbation Theory},
  author={Bender, C.M. and Orszag, S.A.},
  isbn={9780387989310},
  lccn={99044783},
  series={Advanced Mathematical Methods for Scientists and Engineers},
  url={https://books.google.de/books?id=-yQXwhE6iWMC},
  year={1999},
  publisher={Springer}
}

\end{document}